\documentclass[12pt]{elsarticle}

\usepackage{amssymb}
\usepackage{soul}
\usepackage{framed}
\usepackage{graphicx}
\usepackage{float}
\usepackage{url}
\usepackage{rotating}

\usepackage{natbib}
\usepackage{geometry}
\usepackage{fleqn}

\usepackage{epsfig}

\usepackage[table]{xcolor}
\usepackage{hyperref}

\usepackage{pgfplots, pgfplotstable}
\usepackage{scalefnt}
\usepackage{tikz}
\usetikzlibrary{shapes.gates.logic.US,trees,positioning,arrows}
\usetikzlibrary{shapes.multipart}
\usetikzlibrary{shapes,arrows,snakes}

\usepackage{alltt                                    
	, multirow
	,subfig
	, booktabs
	, listings
	, graphicx
	,float
	,cite
	,verbatim
	,mathtools
	,url
	,amsmath
}

\usepackage{verbatim}
\usetikzlibrary{arrows,shapes,backgrounds}
\usepgfplotslibrary{statistics}

\usetikzlibrary{shapes.gates.logic.US,trees,positioning,arrows}
\usetikzlibrary{shapes.multipart}
\usetikzlibrary{shapes,arrows}

\tikzstyle{vertex}=[ellipse,fill=black!25,minimum size=20pt, inner sep=0pt]
\tikzstyle{edge} = [draw,thin,-]
\tikzstyle{glabel} = [text width=1cm,text centered,font=\bf]
\pgfdeclarelayer{bg}    
\pgfsetlayers{bg,main} 

\usepackage{pgfplots, pgfplotstable}
\usepackage{tikz}
\pgfplotsset{compat=1.16}
\usepackage{pgfplotstable}
\usepackage{pgf-pie}

\newif\ifpienumberinlegend
\pgfkeys{/number in legend/.code=
    \expandafter\let\expandafter\ifpienumberinlegend
    \csname if#1\endcsname
    \ifpienumberinlegend

    \def\beforenumber##1\afternumber{}%
    \fi,
    /number in legend/.default=true
}

\makeatletter
\pgfplotsset{
    boxplot/hide outliers/.code={
        \def\pgfplotsplothandlerboxplot@outlier{}%
    }
}
\makeatother

\tikzstyle{vertex}=[ellipse,fill=black!25,minimum size=20pt, inner sep=0pt]
\tikzstyle{edge} = [draw,thin,-]
\tikzstyle{glabel} = [text width=1cm,text centered,font=\bf]
\pgfdeclarelayer{bg}    
\pgfsetlayers{bg,main} 

\usepackage{tcolorbox}

\usepackage{fancybox}
\usepackage{multirow}
\usepackage{flushend}
\usepackage{booktabs}
\usepackage{tabularx}
\usepackage{comment}
\usepackage{array}
\usepackage{mdframed}
\DeclareGraphicsExtensions{.pdf,.png}

\usepackage{subfig}

\usepackage{listings}
\usepackage{courier}
\usepackage{color}
\usepackage{hyphenat}

\usepackage{alltt                                    
	, multirow
	,subfig
	, booktabs
	, listings
	,float
	,verbatim
	,mathtools
	,url
	,amsmath
}
\usepackage{framed,lipsum}
\usepackage{pgfkeys}
\usepackage{graphics} 
\usepackage{graphicx}
\usepackage{pgfplots, pgfplotstable}
\pgfplotsset{compat=1.15}
\usepackage{tikz}
\usetikzlibrary{patterns}

\usepackage{algorithm2e}
\usepackage{algpseudocode}

\newcounter{o}
\usepackage{xspace}

\usepackage{tikz}
\usepackage{pgf-pie}
\usetikzlibrary{positioning,shadows}

\definecolor{1c1}{RGB}{188,162,6}
\definecolor{1c2}{RGB}{137,129,80}
\definecolor{1c3}{RGB}{239,167,31}
\definecolor{1c4}{RGB}{88,194,241}
\definecolor{1c5}{RGB}{6,180,188}

\tikzset{mynode/.style={draw=white,solid,circle,fill=green,inner sep=1pt, thick,
text=black}}
\tikzset{arrow line/.style={dashed, line width= 2.5pt, color=#1}}

\def\bf{\textbf}

\def\it{\textit}

\usepackage{balance}

\normalsize

\usepackage{enumitem}

\usepackage{tikz}

\lstdefinestyle{inlinecode}{basicstyle={\ttfamily\scriptsize\bfseries}}

\newcommand{\urls}[1]{{\scriptsize\url{#1}}}

\usepackage{soul}
\usepackage[outercaption]{sidecap}    

\usepackage{enumitem}
\usepackage[T1]{fontenc}
\usepackage{pifont}

\definecolor{lightgray}{gray}{.92}

\usepackage{graphicx}
\usepackage{subfig}

\definecolor{Gray}{gray}{0.9}

\journal{Journal of Systems and Software}

\begin{document}

\begin{frontmatter}



\title{Reproducibility of Issues Reported in Stack Overflow Questions: Challenges, Impact \& Estimation}


\author{
	Saikat Mondal ~ Banani Roy\\
	Department of Computer Science, University of Saskatchewan, Canada\\
	\{saikat.mondal, banani.roy\}@usask.ca	
}


\begin{abstract}
Software developers often submit questions to technical Q\&A sites like Stack Overflow (SO) to resolve code-level problems. In practice, they include example code snippets with questions to explain the programming issues. Existing research suggests that users attempt to reproduce the reported issues using given code snippets when answering questions.
Unfortunately, such code snippets could not always reproduce the issues due to several unmet challenges that prevent questions from receiving appropriate and prompt solutions. 
One previous study investigated reproducibility challenges discussed in 400 Java questions and produced a catalog of them (e.g., too short code snippets). However, it is unknown how the practitioners (i.e., SO users) perceive this challenge catalog. Practitioners' perspectives are inevitable in validating these challenges and estimating their severity.
In this study, we first surveyed 53 practitioners to understand their perspectives on reproducibility challenges. We attempt to (a) see whether they agree with these challenges, (b) determine the impact of each challenge on answering questions, and (c) identify the need for tools to promote reproducibility.
Survey results show that -- (a) about 90\% of the participants agree with the challenges, (b) ``missing an important part of code'' most severely hurt reproducibility, and (c) participants strongly recommend introducing automated tool support to promote reproducibility.
Second, we extract \emph{nine} code-based features (e.g., LOC, compilability) and build five Machine Learning (ML) models to predict issue reproducibility. Early detection might help users improve code snippets and their reproducibility. 
Our models achieve 84.5\% precision, 83.0\% recall, 82.8\% F1-score, and 82.8\% overall accuracy, which are highly promising.
We also validate the effectiveness of our features by predicting the reproducibility status of C\# code snippets.
Third, we systematically interpret the ML model and explain how code snippets with reproducible issues differ from those with irreproducible issues.

\end{abstract}




\end{frontmatter}

\section{Introduction}
\label{sec:introduction}

Stack Overflow (SO) has emerged as one of the largest and most popular technical Q\&A sites. This Q\&A site is continuously contributing to the body of knowledge in software development \citep{tahaei2020understanding, treude2011programmers, vincent2018examining}. Millions of software developers visit SO daily to solve their programming-related problems \citep{gao2020code2que, datadumpapi}. Around 6K questions are posted in SO every day \citep{datadumpapi}. Among them, many questions discuss code-level problems (e.g., errors, unexpected behavior) \citep{treude2011programmers}. 
In practice, such questions often include example code snippets to explain the code issues. 
Existing research suggests that SO users first attempt to reproduce the issues reported in the questions using the code snippets \citep{Mondal-SOIssueReproducability-MSR2019}. Upon success, they could submit their appropriate solutions.
Reproducibility means a complete agreement between the reported and investigated issues \citep{crashdroid, Mondal-SOIssueReproducability-MSR2019}. Unfortunately, such programming issues could not always be reproduced due to several unmet challenges of the given code snippets \citep{Mondal-SOIssueReproducability-MSR2019, crashscope, soltani2017guided}. 
This phenomenon prevents the questions from getting prompt and appropriate solutions. 
For example, Mondal et al.~ \citep{Mondal-SOIssueReproducability-MSR2019} show that a question with a code snippet capable of reproducing the reported issue has more than three times higher chance of receiving an acceptable answer compared to a question with a code snippet unable to reproduce the issue. Moreover, the median time delay of receiving an accepted answer is double, and the average number of answers is significantly less for questions with irreproducible issues than those with reproducible issues.

Two existing studies \citep{querytousablecode, gistable} investigate the usability (e.g., compilability, executability) challenges of the code snippets posted on Q\&A sites. 
For example, Yang et al.~\citep{querytousablecode} analyzes the parsability and compilability of code snippets extracted from the accepted answers of SO. However, their fully automatic analysis only exposed parse and compile errors (e.g., syntax errors,  incompatible types). 
Horton and Parnin~\citep{gistable} examine the executability of the Python code snippets found on the GitHub Gist system. They identify several flaws (e.g., syntax errors, indentation errors) that prevent the executability of such code snippets. However, a simple execution success does not always guarantee the reproducibility of issues \citep{Mondal-SOIssueReproducability-MSR2019}. 
Several studies investigate the quality of SO code snippets by measuring their readability \citep{buse2008metric, buse2009learning, daka2015modeling, posnett2011simpler, scalabrino2016improving, treude2017understanding}, and understandability \citep{lin2008evaluation, scalabrino2017automatically, trockman2018automatically}. Unfortunately, their capability of reproducing the issues reported in SO questions was not investigated. 
Mondal et al.~\citep{Mondal-SOIssueReproducability-MSR2019} first investigate the reproducibility of issues reported on 400 SO questions related to Java programming language. Their investigation produces a catalog of challenges (see Table \ref{table:reproducibility-challenges}) that might prevent reproducibility. However, the catalog was not validated by the practitioners (i.e., SO users). Thus, it is unknown how practitioners perceive these reproducibility challenges. 
The practitioners' perspective is essential for (a) estimating the validity and severity of these challenges and (b) introducing effective tool supports to promote reproducibility.

\begin{table}[htb]
	\centering
	\caption{Challenges preventing issue reproduction}
	\label{table:reproducibility-challenges}
    \resizebox{3in}{!}{%
    \begin{tabular}{p{8cm}}  \toprule

    (C\textsubscript{1}) Class/Interface/Method not found \\ 
    (C\textsubscript{2}) Important part of code missing \\
    (C\textsubscript{3}) External library not found \\
    (C\textsubscript{4}) Identifier/Object type not found \\
    (C\textsubscript{5}) Too short code snippet \\
    (C\textsubscript{6}) Database/File/UI dependency \\
    (C\textsubscript{7}) Outdated code  \\ \bottomrule
    
    \end{tabular}
    }
\end{table}

In this study, we first survey $53$ users of SO 
to understand their perspective on reproducibility challenges (Table \ref{table:reproducibility-challenges}).
This survey investigates three perspectives -- (a) agreement (agree/disagree) with the reproducibility challenges, (b) estimation of the potential impact of these challenges, and (c) determination and prioritization of tool support needs. Second, we develop five Machine Learning (ML) models to predict issue reproducibility. Nine code-based features (e.g., LOC, compilability) are extracted to develop these models. Their strength and generalizability are also verified. Third, we systematically interpret the ML model to explain how the code snippets with reproducible issues differ from those with irreproducible issues. 
Such interpretation might help users improve their code snippets to support reproducibility.

\vspace{2mm}
\noindent \textbf{Contribution.} 
This paper significantly extended several aspects of our previous study (Mondal et al.~\citep{mondal2022reproducibility}).
\emph{First,} we attempt to see how well empirical evidence supports practitioners' perspectives on the impact of reproducibility challenges in answering questions.
\emph{Second,} we extract nine code-related features to estimate code snippets' capability of reproducing the issues reported in the questions.
\emph{Third,} we developed five ML models using these features to classify reproducible and irreproducible issues.
\emph{Fourth,} we systematically interpret the results of the ML model to distinguish between code snippets with reproducible and irreproducible issues.
\emph{Fifth,} we prepare a dataset by manually investigating the reproducibility of issues reported in 100 C\# related questions. We use this dataset to examine the effectiveness of our extracted nine features by predicting the reproducibility status of C\# code snippets.

\vspace{2mm}
\noindent\textbf{Replication Package} that contains the survey questionnaire, all the responses \& their analysis, and our prepared C\# dataset is shared in our online appendix \citep{ourdataset}.

\section{Background and Related Work}
\label{sec:background}
In this section, we start by introducing the \emph{issue reproducibility} and the associated challenges that prevent reproducibility. We then proceed to describe related studies. 

\begin{figure}[!htb]
\centering
  \includegraphics[width=5in]{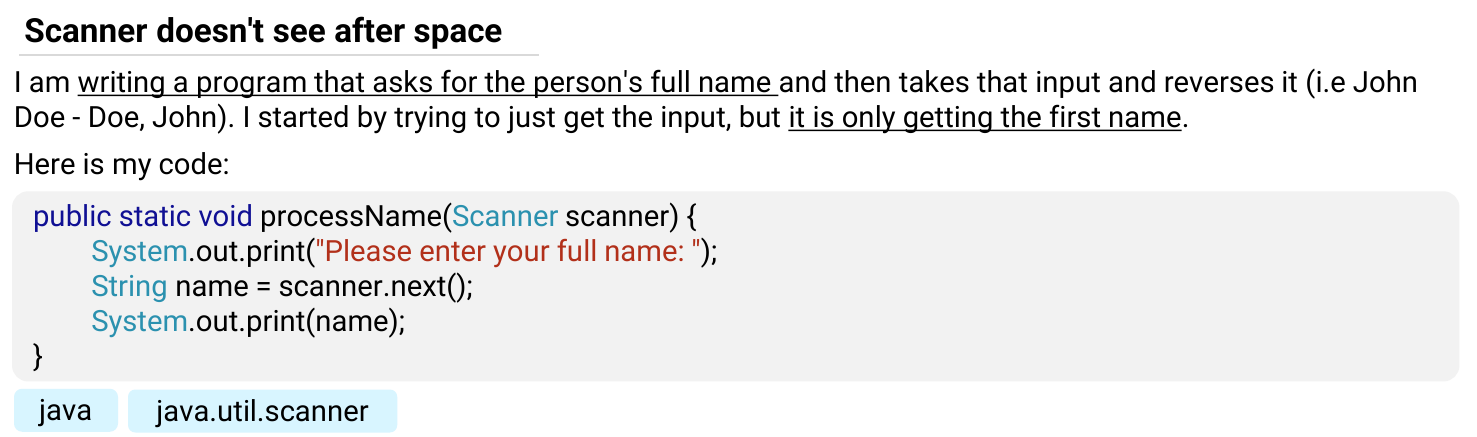}
  \caption{An example \citep{footnote1} question of Stack Overflow that discusses a programming issue.}
  \label{fig:example-issue-reproducibility}
\end{figure}

\subsection{Issue Reproducibility}

Reproducibility is often closely related to repeatability and replicability \citep{anda2008variability}. However, its definition differs across disciplines. 
In this study, reproducibility refers to a complete agreement between the reported and the investigated issues \citep{crashdroid, Mondal-SOIssueReproducability-MSR2019}. 
Consider the example question in Fig. \ref{fig:example-issue-reproducibility}, where a user was trying to take a person's full name (e.g., John Doe) as input by invoking the \texttt{next()} method of Java \texttt{Scanner} class. 
However, when the user printed the name, only the first name (e.g., John) was displayed, and the last name (e.g., Doe) was missed. Specifically, the user did not retrieve the part of the name following the space.
She included the definition of the method \texttt{processName()}, where she was taking and printing the name. 
According to our definition, the issue is reproducible when users get only the first part of the name and lose the part after space by invoking the \texttt{processName()} method. 
According to our definition, the issue is reproducible when users invoke the \texttt{processName()} method, can retrieve only the first part of the name, and lose the part after space.

\subsection{Reproducibility Challenges}

Mondal et al.~\citep{Mondal-SOIssueReproducability-MSR2019} investigated the reproducibility of issues reported in 400 Java-related questions of SO. 
They could not reproduce about 22\% issues due to several unmet challenges of the code snippets, as shown in Table \ref{table:reproducibility-challenges}. 
This section briefly discusses the challenges that prevent reproducibility as follows.

\begin{figure}[htb]
\centering
  \includegraphics[width=5.5in]{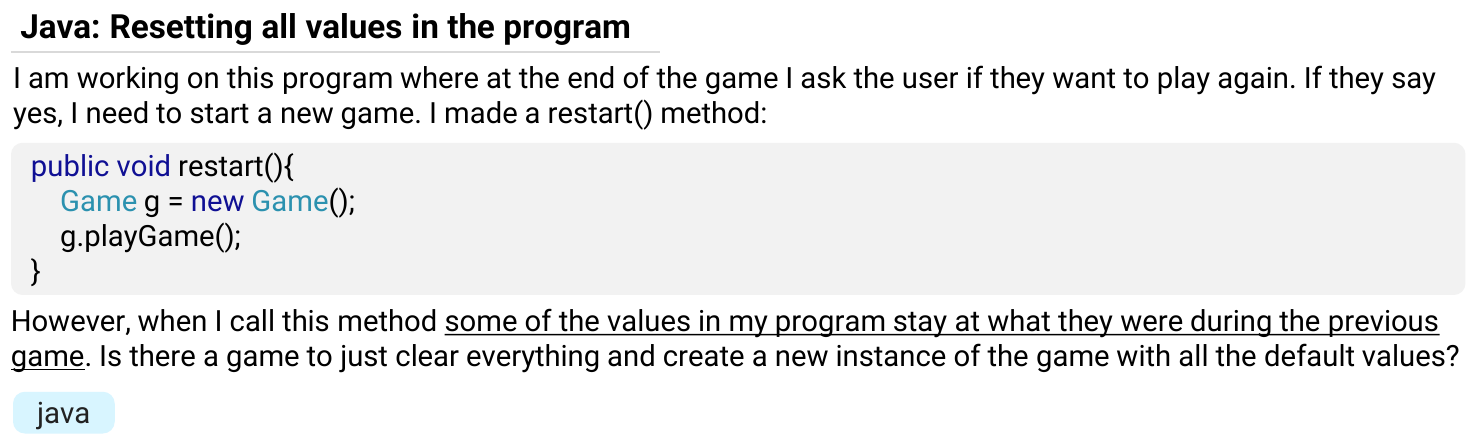}
  \caption{An example \citep{footnote2} question of Stack Overflow whose issue could not be reproduced due to mainly two unmet challenges -- (i) class/interface/method not found and (ii) important part of code missing.}
  \label{fig:reproducibility-challenge-1}
\end{figure}

    (C\textsubscript{1}) \textbf{Class/Interface/Method not found.} Developers often submit only the code snippets of interest with questions that are neither complete nor compilable. These snippets may lack class definitions or invoke methods from other classes without providing the method definitions.
    Fig. \ref{fig:reproducibility-challenge-1} shows an example question where the developer attempts to reset all the variables to their default values. 
    However, the code did not work as expected, and some variables retained their old values. Unfortunately, this issue could not be reproduced due to missing elements, particularly definitions of the \texttt{Game} class and \texttt{playGame()} method. These are essential to reproduce the issue. Thus, while attempting to answer the question, one developer commented – \emph{``Can you post more code? The Game class? The class that contains the restart() method?''}.
    
    (C\textsubscript{2}) \textbf{Important part of code missing.} Code snippets often lack essential code statements, making it challenging to reproduce issues. "An important part" refers to a part of code that cannot be accurately inferred or guessed. For example, adding definitions for the \texttt{Game} class and \texttt{playGame()} method in Fig. \ref{fig:reproducibility-challenge-1} might make the code executable. However, we cannot reproduce the issue. We require the specific code within the class and the method used by the developer to reset the variables.
    
\begin{figure}[htb]
  \centering
  \includegraphics[width=5.5in]{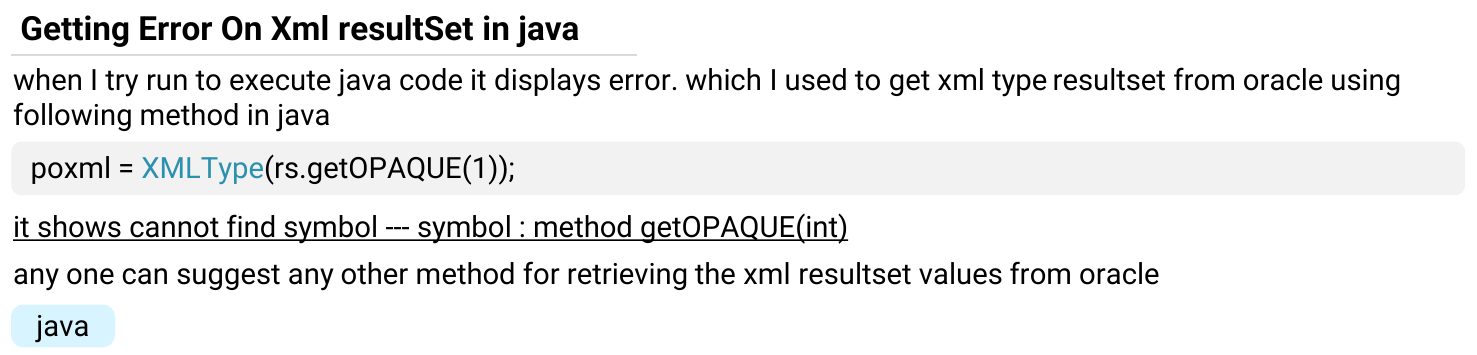}
  \caption{An example \citep{footnote3} question of Stack Overflow whose issue could not be reproduced due to mainly three unmet challenges -- (i) external library not found, (ii) identifier/object type not found, and (iii) too short code snippet.}
  \label{fig:reproducibility-challenge-2}
\end{figure}
    
    (C\textsubscript{3}) \textbf{External library not found.} Resolving external library dependencies is one of the major challenges in reproducing the issues from the submitted code snippets~\citep{Mondal-SOIssueReproducability-MSR2019}. Java has thousands of external (i.e., third-party) libraries with millions of classes and methods. Different libraries contain classes and methods with the same name. Thus, developers face difficulty adding the appropriate libraries if the code snippets miss import statements of the external libraries or question descriptions do not have hints that point to the appropriate libraries. As a result, they could not compile/execute the code and failed to reproduce the question issues. In the code example shown in Fig. \ref{fig:reproducibility-challenge-2}, the import statement for the library containing the \texttt{XMLType} class was missing. Hence, the issue could not be reproduced.
    
    (C\textsubscript{4}) \textbf{Identifier/Object type not found.} Code snippets often use identifiers/objects without declaring them. Sometimes, developers can infer the types of these identifiers/objects by examining assigned values or method invocations. However, when the types are not easily inferred, it can hinder issue reproducibility. Consider the code example in Fig. \ref{fig:reproducibility-challenge-2}, where the types of \texttt{poxml} and \texttt{rs} are unknown and hard to guess from the submitted code.
    
    (C\textsubscript{5}) \textbf{Too short code snippet.} Developers often submit incomplete, too-short code snippets with their questions, making it difficult to reproduce the reported issues. For example, Fig. \ref{fig:reproducibility-challenge-2} shows a question where only \emph{one} line of code was provided. Thus, guessing the missing statements and making the code compilable/executable to reproduce the issue is challenging. The ``too short code snippet'' challenge could overlap with ``important parts of code missing''. However, even a longer code could miss important code segments.
    
    \begin{figure}[htb]
      \centering
      \includegraphics[width=5.5in]{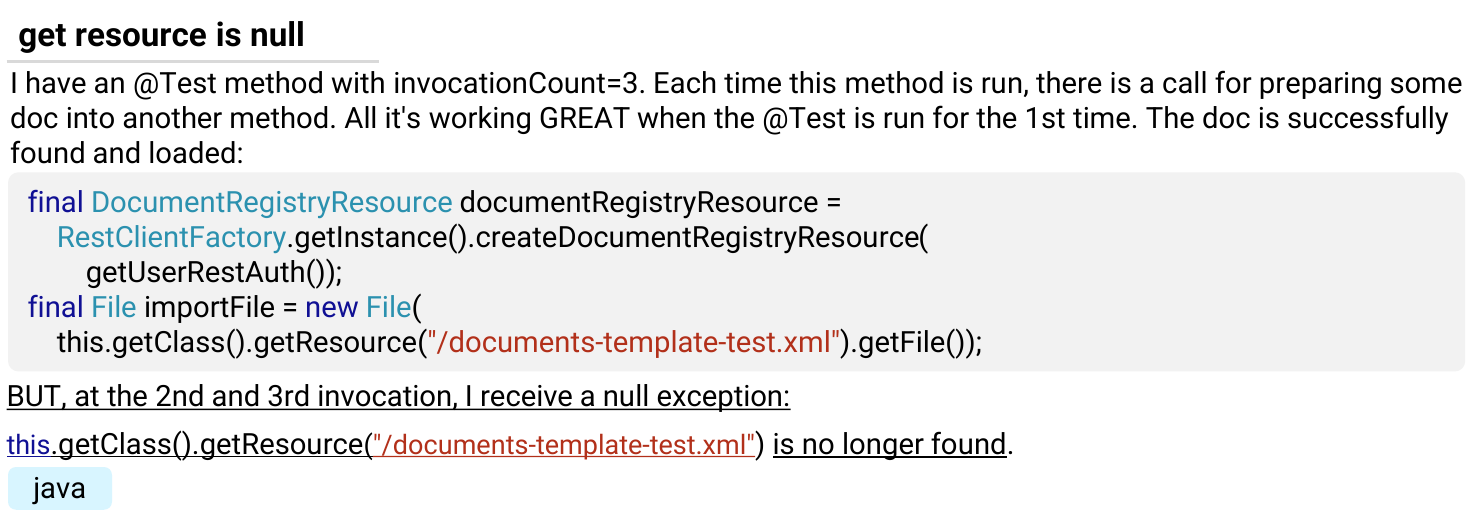}
      \caption{\small{An example \citep{footnote4} question of Stack Overflow whose issue could not be reproduced due to mainly two unmet challenges -- (i) database/file/UI dependency and (ii) class/interface/method not found.}}
      \label{fig:reproducibility-challenge-3}
    \end{figure}
        
    \begin{figure}[htb]
      \centering
      \includegraphics[width=5.5in]{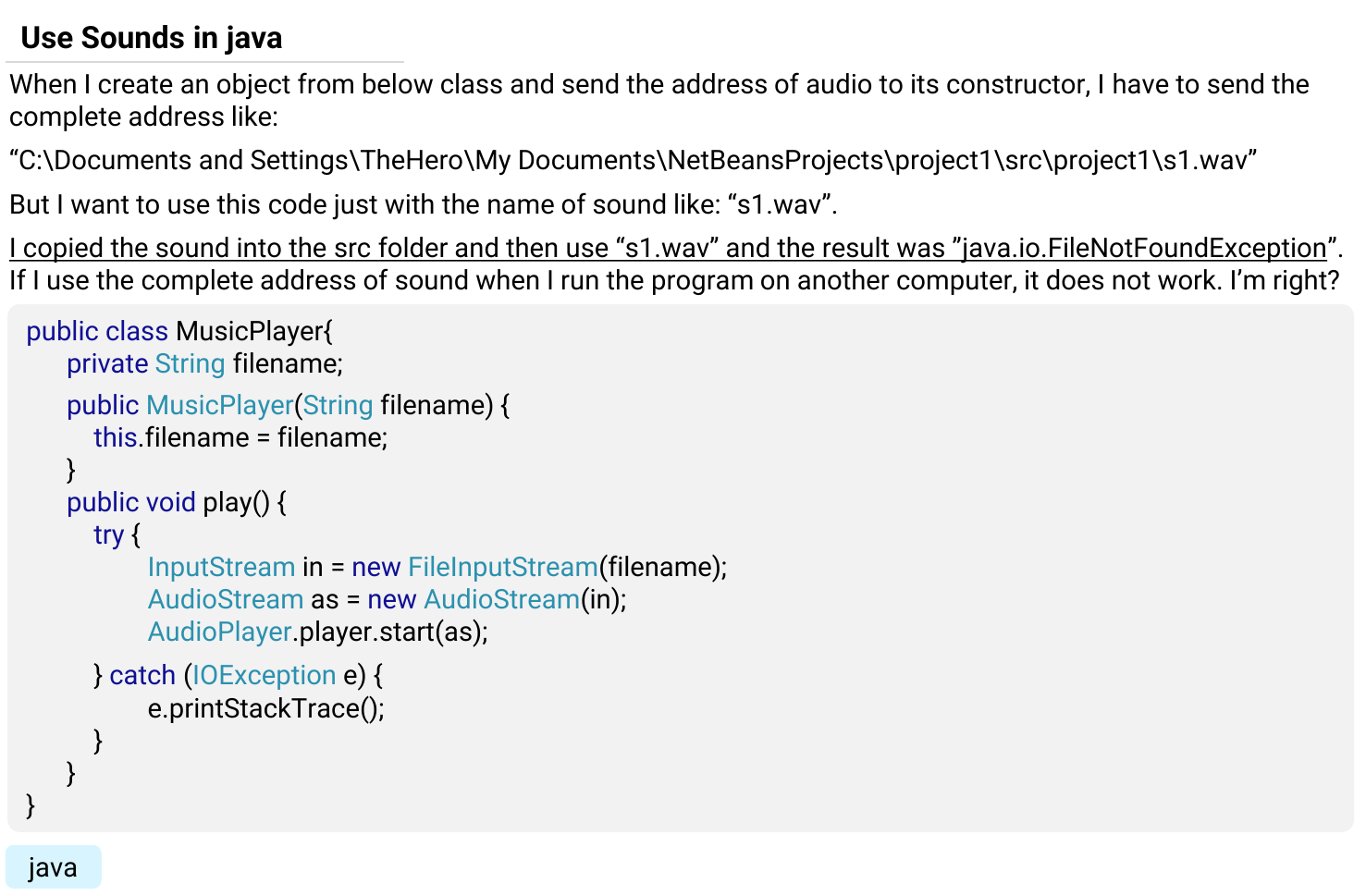}
      \caption{An example \citep{footnote5} question of Stack Overflow whose issue could not be reproduced due to mainly outdated code challenge.}
      \label{fig:reproducibility-challenge-4}
    \end{figure}
    
    (C\textsubscript{6}) \textbf{Database/File/UI dependency.} Several code snippets could not reproduce issues due to their complex interactions with databases, external files, and UI elements. Fig. \ref{fig:reproducibility-challenge-3} shows an example question where the developer tried to create and load an XML doc but encountered a null exception when attempting to access the doc multiple times. This issue could not be reproduced due to the external file dependency. The XML document could not be created because the question lacks crucial (e.g., file path, content) associated with the file.
    
    (C\textsubscript{7}) \textbf{Outdated code.} A few code snippets contain outdated code (e.g., deprecated class/API) that could prevent issue reproducibility. In this study, we consider a code outdated when it contains classes/APIs incompatible with JDK-1.8. Sometimes, finding an equivalent or alternative class/API for the outdated ones can be challenging. For example, in Fig. \ref{fig:reproducibility-challenge-4}, the \texttt{AudioStream} and \texttt{AudioPlayer} classes are not used in JDK-1.8, making it challenging for developers to find alternatives and reproduce the issue.

\subsection{Related Work}
\label{relatedwork}

Yang et al.~\citep{querytousablecode} analyze the usability (e.g., parsability, compilability) of about 914K Java code snippets extracted from the accepted answers of SO. They expose several  parsability (e.g., syntax error) and compilability (e.g., incompatible types) challenges. 
The authors employ automated tools like Eclipse JDT and ASTParser to parse and compile code snippets. Their analysis identifies challenges that prevent the code snippets from parsing and compiling.
Horton and Parnin~\citep{gistable} investigate the executability of Python code found on the GitHub Gist system. They report the types of execution failures encountered while running Python gists, such as import, syntax, and indentation errors. However, the success of a code snippet's execution does not always guarantee the reproducibility of an issue. 
A few challenges of compilability/executability might overlap with the reproducibility challenges. However, reproducibility requires testing and debugging that warrant manual analysis, which was not done by Horton and Parnin~\citep{gistable}.  

Mondal et al. \citep{Mondal-SOIssueReproducability-MSR2019} go beyond code execution by manually investigating the reproducibility of issues reported in 400 Java questions using the included code snippets. They produce a catalog of challenges (e.g., an important part of code missing) that prevent reproducibility. However, listening to the practitioners to determine whether they agree with these challenges is essential. Furthermore, understanding the potential impact of these challenges in answering questions is also important, which was not addressed by any earlier studies.

Several researchers investigate the challenges of reproducing software bugs and security vulnerabilities \citep{erfani2014works, rahman2020why, mu2018understanding}. 
Joorabchi et al. \citep{erfani2014works} analyze 1,643 non-reproducible bug reports to find causes of their non-reproducibility and reveal six root causes (e.g., environmental differences, insufficient information).
Rahman et al.~\citep{rahman2020why} also investigate 576 non-reproducible bug reports from two popular software systems (e.g., Firefox and Eclipse). They identify 11 challenges (e.g., bug duplication, missing information, ambiguous specifications) that might prevent bug reproducibility. They surveyed 13 developers to understand how they handle non-reproducible bugs. They found that developers either close these bugs or solicit further information.
Mu et al.~\citep{mu2018understanding} analyze 368 security vulnerabilities. Their study suggests that individual vulnerability reports often lack sufficient information (e.g., configurations, OS) to reproduce the reported vulnerabilities.
Their survey of experts in software security suggested that, besides internet-scale crowd-sourcing and some interesting heuristics, manual efforts (e.g., debugging) based on experience are the sole way to retrieve missing information from reports. 
Our study is similar in methodology and problem aspects. However, our research context differs from theirs since we study how practitioners perceive reproducibility challenges of issues reported in SO questions.

Tahaei et al.~\citep{tahaei2020understanding} analyze 1,733 privacy-related SO questions to understand developers' challenges and confusion while dealing with privacy-related topics.
Ebert et al.~\citep{ebert2019confusion} investigate the reasons (e.g., missing rationale) and impacts (e.g., merge decision is delayed) of confusion in code reviews. Their study shows how developers handle confusion in code reviews, such as asking for more information, improving their understanding of existing code, and discussing issues offline. They surveyed developers to gain insights for researchers and tool developers. While our study uses a similar method, our research goals are different. We surveyed 53 developers to understand how reproducibility challenges impact SO answers and identify tool requirements to address these challenges to promote reproducibility.

Ford et al.~\citep{ford2018we} implement a month-long, just-in-time mentorship program for SO. Mentors guide the novices with informative feedback on their questions, aiming to reduce delays in getting answers or adverse feedback. However, human mentorship is costly and challenging to sustain. Horton and Parnin~\citep{horton2019dockerizeme} develop DockerizeMe, a system that infers dependencies necessary to execute Python code snippets without import errors. Terragni et al.~\citep{terragni2016csnippex} propose CSnippEx, which automatically converts Java code snippets into compilable Java source code files. 
We seek the practitioners' recommendations on the design requirement to introduce an automatic tool to support reproducibility. As an initial step, we extract nine code-related features and develop five ML models to predict the reproducibility (reproducible/irreproducible) of issues.

\section{Study Design}
\label{sec:study-design}

We divided the study into two phases: (1) we survey practitioners for their perspectives on reproducibility challenges, and (2) we predict the reproducibility status using ML models.

\subsection{Practitioners’ Perspectives on Reproducibility Challenges}

In this phase, we survey 53 practitioners (i.e., SO users) to understand their perspectives on reproducibility challenges and estimate the impacts of these challenges on answering questions. We also seek practitioners' suggestions on the tool's design to support reproducibility. In this phase, we answer five research questions.

\smallskip
\textbf{RQ\textsubscript{1}:What challenges do practitioners perceive as contributing to the lack of reproducibility of issues reported on Stack Overflow questions?}
Mondal et al.~\citep{Mondal-SOIssueReproducability-MSR2019} produce a reproducibility challenge catalog (refer to Table \ref{table:reproducibility-challenges}). Nonetheless, developers' feedback on such empirical findings is essential to increase confidence in these findings \citep{rahman2020why}. 
To answer RQ\textsubscript{1}, we present four example SO questions (issue descriptions + code snippets). Participants were asked to reproduce the reported issues using the provided code snippets. 
Upon failure, we ask for their agreement with the listed reproducibility challenges by Mondal et al.~\citep{Mondal-SOIssueReproducability-MSR2019}. 
Fifty-three software developers validated these challenges, with an agreement level ranging from 80\% to 94\%. Participants also report three additional challenges they encountered.

\textbf{RQ\textsubscript{2}: What are the perceived impacts of the reproducibility challenges in answering a Stack Overflow question?}
Understanding the impact of each challenge is important to determine which ones must be addressed to receive appropriate solutions. 
To answer RQ\textsubscript{2}, we present each challenge with five options -- (1) \emph{not a problem}, (2) \emph{moderate}, (3) \emph{severe}, (4) \emph{blocker}, and (5) \emph{no opinion}. 
We then ask participants to select one of these options. Most participants assessed ``an important part of code missing'' as a blocker and ``outdated code'' as not a problem.

\textbf{RQ\textsubscript{3}: Do empirical findings support the practitioners' perspective on the impacts of the reproducibility challenges?}
Examining empirical findings against practitioners' perspectives on reproducibility challenges ensures practical and impactful insights.
We categorize the questions with irreproducible issues based on their specific challenges. Then, we determine the correlation between question category and question meta-data. Our analysis shows that \emph{important part of code missing}, \emph{too short code snippet}, and \emph{external library not found} prevent questions from getting answers, including acceptable ones. This evidence aligns with practitioners' perspectives on the impacts of reproducibility challenges.

\textbf{RQ\textsubscript{4}: How do developers prioritize to address the reproducibility challenges?}
Prioritizing challenges is essential to identify those that need to be fixed within tight constraints (e.g., time or budget). Participants were asked to prioritize three reproducibility challenges. The top three priorities identified are -- (i) an important part of code missing, (ii) class/interface/method not found, and (iii) too short code snippet.

\textbf{RQ\textsubscript{5}: What are the interactive tool design requirements to address the reproducibility challenges?}
Questions with reproducible issues have a significantly higher chance of receiving timely answers \citep{Mondal-SOIssueReproducability-MSR2019}. However, SO's current question submission system cannot address reproducibility issues. Aiming to introduce tool support, we ask practitioners for tool design requirements to ensure practical solutions to promote reproducibility. They suggest introducing browser/IDE plugins to assist question submitters in improving their code snippets to support reproducibility. They also recommended tools to identify (1) missing important parts of code and (2) reproducibility challenges according to their severity by statically analyzing the code snippets.


\subsection{Prediction of Reproducibility Status}

In this phase, we develop ML models by extracting code-based features to predict whether code snippets included in questions can reproduce the reported issue. We answer the following research question.

\textbf{RQ\textsubscript{6}: Can we predict the reproducibility status of the issues reported in Stack Overflow questions?}
We extract nine code-based features by analyzing two types of code snippets -- \emph{can reproduce} \& \emph{cannot reproduce} question issues. We developed five supervised ML models using these features to classify reproducible issues from irreproducible ones. Our models can predict issue reproducibility with 84.5\% precision, 83.0\% recall, 82.8\% F1-score, and 82.8\% overall accuracy. To validate our extracted features, we tested two machine learning models to predict issue reproducibility reported in 100 C\# related questions and found promising results.

\section{Practitioners’ Perspectives on Reproducibility Challenges (RQ\textsubscript{1} -- RQ\textsubscript{5})}
\label{sec:practitioners-perspectives}

In this section, we first survey practitioners to understand \emph{``what they say''} (\textbf{RQ\textsubscript{1}}, \textbf{RQ\textsubscript{2}}, and \textbf{RQ\textsubscript{3}}) about reproducibility challenges and the impact of these challenges to answer questions. We then investigate whether empirical evidence supports practitioners' opinions (\textbf{RQ\textsubscript{4}}).
Next, we see the code modification plan to understand \emph{``what developers do''} (\textbf{RQ\textsubscript{4}}) to reproduce the issues. 
Then, we find \emph{``what practitioners suggest''} (\textbf{RQ\textsubscript{5}}) to get recommendations for designing interactive tools to promote reproducibility.
We utilize appropriate tools to analyze the survey responses and answer our research questions.

\subsection{Survey}
\label{subsec:survey}

We conduct an online survey to validate and expand the catalog of reproducibility challenges (see Table \ref{table:reproducibility-challenges}) identified by Mondal et al.~\citep{Mondal-SOIssueReproducability-MSR2019}. Furthermore, we seek participants' viewpoints on the impacts of these challenges and their potential fixes. We follow Kitchenham and Pfleeger's guidelines for personal opinion surveys \citep{kitchenham2008personal}. However, we also consider the guidance and ethical issues from the established best practices~\citep{groves2011survey, singer2002ethical}.

\smallskip
\textbf{Survey Design.}
Our survey includes different questions (e.g., multiple-choice and free-text answers). Before asking questions, we explain the survey's purpose and our research goals. We assure the participants that their information must be treated confidentially. First, we piloted the survey with three practitioners to get feedback on its length and clarity. Based on their input, we made minor adjustments and finalized the survey. We informed participants that the survey would take 45-50 minutes based on the pilot survey. Responses from the pilot survey are excluded from the results presented in this paper. Our survey comprises five parts as follows.


    (i) \emph{Consent and Prerequisite.} In this part, we ask the participants to confirm their consent to participate in this survey and agree to process their data. We set two constraints for participating in this survey -- (1) participants must answer SO questions and (2) have experience in Object-Oriented Programming (OOP), preferably Java.
    
    (ii) \emph{Participants Information.} This part collects participants' OOP and professional software development experience, professional background, SO account age, and question-answering experience.
    
    (iii) \emph{Agreement to the Reproducibility Challenges.} We present four SO questions (see Fig. \ref{fig:reproducibility-challenge-1}, \ref{fig:reproducibility-challenge-2}, \ref{fig:reproducibility-challenge-3} \& \ref{fig:reproducibility-challenge-4}) from Mondal et al.'s \citep{Mondal-SOIssueReproducability-MSR2019} published dataset. 
    We aim to select the fewest questions that cover all reproducibility challenges and minimize common challenges among questions.
    Our selected questions share only one common challenge (i.e., Class/Interface/Method not found) between Q1 (Fig. \ref{fig:reproducibility-challenge-1}) \& Q3 (Fig. \ref{fig:reproducibility-challenge-3}). Our primary focus is to validate the challenge catalog. We thus ask participants to reproduce the question issues using the provided code snippets. Upon selection, ``I cannot reproduce'', we ask for their agreement with the provided challenges that could prevent reproducibility. However, they have the option to report additional challenges. 
    We ask the following three questions and analyze participants' agreement with Mondal et al.'s \citep{Mondal-SOIssueReproducability-MSR2019} findings on reproducibility status and challenges.

        \textbf{Q\textsubscript{1})} Do the participants reproduce the reported issues? (\emph{I can reproduce the issue/I cannot reproduce the issue})
        
        \textbf{Q\textsubscript{2})} Do the participants agree with the reproducibility challenges? (\emph{Yes/No})
        
        \textbf{Q\textsubscript{3})} Do the participants find additional challenges reproducing the issues? (\emph{Text})
        
    \noindent On the contrary, upon selection, ``I can reproduce'', we ask for their code to see the editing actions needed to reproduce the issues.
    
    (iv) \emph{Impact and severity of the Reproducibility Challenges.} Here, each challenge is presented with five options: (1) \emph{not a problem}: no difficulties in answering; (2) \emph{moderate}: irritating but answerable appropriately; (3) \emph{severe}: time-consuming but answerable; (4) \emph{blocker}: unable to reproduce the issue and thus unable to answer; (5) \emph{no opinion}: no specific view. In particular, we ask the question to the participants as follows.

        What are the impacts of the reproducibility challenges in answering a question? (\emph{not a problem/moderate/severe/blocker/no opinion})

    \noindent Participants were asked to prioritize the three most severe reproducibility challenges by answering the following question.

        Which three challenges the participants would like to prioritize above others? (\emph{first choice/second choice/third choice})
    
    (v) \emph{Tool Support Needs.} Finally, we gather participants' opinions on an intelligent tool's needs and design requirements to promote reproducibility. We also offer six tool support options (Table \ref{table:tool-support-options}) and employ a 5-point Likert Scale \citep{joshi2015likert, vagias2006likert} to estimate the participants' consent. In particular, we ask two questions as follows.

        \textbf{Q\textsubscript{1})} What kind of tool support do the participants need to assist with the reproducibility challenges? (\emph{Text})
        
        \textbf{Q\textsubscript{2})} How do the participants agree with our tool support options? (\emph{see options from Table \ref{table:tool-support-options}})

\smallskip
\textbf{Recruitment of Survey Participants.} We recruit participants who provide their consent through the following two ways.

    (i) \emph{Snowball Approach:} We reach out to participants from companies worldwide through personal contacts. Using a snowballing method \citep{bi2021accessibility}, we encourage them to invite colleagues with similar experiences to join our survey. Out of 48 participants who consented, 44 were allowed to complete the survey. The remaining four did not meet our participation constraints.
    
    (ii) \emph{Open Circular:} We find participants by posting our study description and goals in specialized Facebook groups where software developers discuss programming issues. We also use LinkedIn, a large professional network, to reach potential participants. Through this open circular, we recruited 15 participants who met our criteria and finally received nine valid responses.

\begin{figure}[!htb]
	\centering
	\subfloat[Java experience (in years)]
    {
	\resizebox{2.7in}{!}{
    \begin{tikzpicture}
    \pie[explode=0.1, text=pin, number in legend, sum = auto, color={black!0, black!15, black!30, black!45, black!60, black!75}]
        { 23/\Large{$\leq$ 2 (43.4\%)},
          11/\Large{3--5 (20.7\%)},
          8/\Large{6--8 (15.1\%)},
          6/\Large{9--11 (11.3\%)},
          2/\Large{12--14 (3.8\%)},
          3/\Large{$\geq$ 15 (5.7\%)}
        }
    \end{tikzpicture}
    \label{fig:java-experience}
    }
    }
    \subfloat[Profession]
    {
	\resizebox{2.2in}{!}{
    \begin{tikzpicture}
    \pie[explode=0.1, text=pin, number in legend, sum = auto, color={black!0, black!20, black!40, black!60}]
        { 24/\Large{SD (45.3\%)},
          4/\Large{TL (7.6\%)},
          5/\Large{RE (9.4\%)},
          20/\Large{AP (37.7\%)}
          }
    \end{tikzpicture}
    \label{fig:profession}
    }
    }
\caption{Information of the survey participants \footnotesize{(\textbf{SD:} Software Developer, \textbf{TL:} Technical Lead, \textbf{RE:} Research Engineer, \textbf{AP:} Academic Practitioner)}.}
\label{fig:info-survey-participants}
\end{figure}
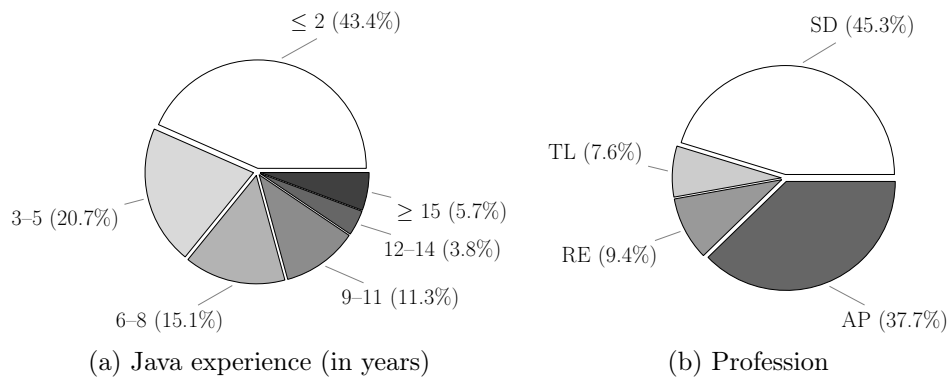

Fig. \ref{fig:info-survey-participants} summarizes the Java experience and professions. Out of the 53 participants, 34 (about 64\%) had five years or less of Java experience (see Fig. \ref{fig:java-experience}). Less experienced developers tend to be more active on SO. Most participants were software developers (45.3\%) or academic practitioners (37.7\%) (e.g., faculty members, students, and postdoctoral researchers) (see Fig \ref{fig:profession}). However, four technical leads and five research engineers also participated. Other details of the participants can be found in our online appendix \citep{ourdataset}.

\textbf{Survey Data Analysis.}
We received 53 valid survey responses and analyzed them using appropriate tools and techniques. For multiple-choice questions, we calculated the percentage for each option chosen. Likert-scale ratings were used to estimate agreement levels. Reproducibility challenges were ranked using the Borda count method \citep{Yamashita-DoDevelopersCareAboutCodeSmell-WCRE2013}. We also manually investigate participants' comments on the impacts of these challenges and their tool design requirements.

\subsection{Agreement Analysis to the catalog of Reproducibility Challenges (RQ1)} \label{agree-analysis-rq1}

\begin{table}[!htb]
\centering
	\caption{Developers agreement to the reproducibility status and the challenges that prevent reproducibility}
	\label{table:results-agreement-analysis}
 	\resizebox{5.8in}{!}{%
    \begin{tabular}{l|c|c|c|c|c} \toprule
    \multirow{2}{*}{\textbf{Question No.}} & \multicolumn{2}{c|}{\textbf{Reproducibility Status}} & \multirow{2}{*}{\textbf{Reproducibility Challenges}} & \multicolumn{2}{c}{\textbf{Reproducibility Challenges}} \\ 
                       & \textbf{Irreproducible} & \textbf{Reproducible} &                 & \hspace{7mm}\textbf{Agree}\hspace{7mm} & \textbf{Disagree}    \\    \midrule
                     \multirow{2}{*} {01 (see Fig. \ref{fig:reproducibility-challenge-1})}  &  \multirow{2}{*} {94.3\%} &   \multirow{2}{*} {5.7\%}  &  \multicolumn{1}{l|}{$\bullet$ Class/Interface/Method not found}  &  92.0\%   &  8.0\%    \\
                                                                                            &                                 &                       & \multicolumn{1}{l|}{$\bullet$ Important part of code missing}     &  92.0\%   &  8.0\%  \\ \midrule
                     
                     \multirow{3}{*} {02 (see Fig. \ref{fig:reproducibility-challenge-2})}  &  \multirow{3}{*} {92.5\%}       &  \multirow{3}{*} {7.5\%}  &  \multicolumn{1}{l|}{$\bullet$ External library not found}  &  93.9\%   &  6.1\%   \\
                                                                                            &                                 &                       & \multicolumn{1}{l|}{$\bullet$ Identifier/Object type not found}     &   89.8\%   & 10.2\%    \\ 
                                                                                            &                                 &                       & \multicolumn{1}{l|}{$\bullet$ Too short code snippet}     &    93.9\%    & 6.1\%  \\ \midrule

                     \multirow{2}{*} {03 (see Fig. \ref{fig:reproducibility-challenge-3})} &  \multirow{2}{*} {92.5\%}        &   \multirow{2}{*} {7.5\%}  &  \multicolumn{1}{l|}{$\bullet$ Class/Interface/Method not found}  &  81.6\%   &  18.4\%  \\
                                                                                           &                                  &                            & \multicolumn{1}{l|}{$\bullet$ Database/File/UI dependency}      &  89.8\%   & 10.2\%    \\ \midrule

                     04 (see Fig. \ref{fig:reproducibility-challenge-4})                   &               73.6\%              &          26.4\%          &   \multicolumn{1}{l|}{$\bullet$ Outdated code} &    79.5\%     & 20.5\%    \\ \midrule
                    \textbf{ Overall}                                                      &               \textbf{ 88.2\%}    &    \textbf{11.8\%}        &                                --           &    \textbf{89.1\%}     &  \textbf{10.9\%}  \\ \bottomrule 
                     
    \end{tabular}
    }
\end{table}

Table \ref{table:results-agreement-analysis} shows the agreement summary. For question 1 (Fig. \ref{fig:reproducibility-challenge-1}), 94.3\% of the participants could not reproduce the issue, while 5.7\% reported successful reproduction. Two reproducibility challenges were identified for this code example: \emph{Class/Interface/Method not found} and \emph{Important part of code missing}. 
These challenges were encountered by 92\% of participants, with 8\% disagreeing. 
A similar pattern of agreement and disagreement regarding reproducibility status and associated challenges were observed for questions 2 (Fig. \ref{fig:reproducibility-challenge-2}) and 3 (Fig. \ref{fig:reproducibility-challenge-3}). 
For question 4 (Fig. \ref{fig:reproducibility-challenge-4}), about 26\% of developers were able to reproduce the issue, where the challenge was \emph{outdated code}. 
The code snippet included classes (e.g., \texttt{AudioStream}) that are not compatible with JDK-1.8. Some developers could reproduce the issue using previous JDK versions or alternative classes. Participants were asked to submit modified code snippets. Some developers used alternative classes, and a few submitted codes were different from the code provided in the original question.

\begin{table}[!htb]
	\centering
	\captionsetup{justification=centering, labelsep=newline}
	\caption{Agreement analysis by experience and profession (\textbf{SD:} Software Developer, \textbf{TL:} Technical Lead, \textbf{RE:} Research Engineer, \textbf{AP:} Academic Practitioner)}
	\label{table:agreement-experience-profession}
	\resizebox{5in}{!}{%
    \begin{tabular}{l|c|c|c|c} \toprule
    \multicolumn{5}{c}{\textbf{Analysis by Java experience}} \\ \midrule
    \multirow{2}{*}{\textbf{Experience}} & \multicolumn{2}{c|}{\textbf{Reproducibility Status}} & \multicolumn{2}{c}{\textbf{Reproducibility Challenges}} \\ 
                                    & \textbf{Irreproducible}   & \textbf{Reproducible} & \textbf{Agree}   & \textbf{Disagree}\\ \midrule
    $\leq$ 5 years & 91.2\%  & 8.8\% & 85.1\%  & 14.9\%   \\ \midrule
    $>$ 5 years & 82.9\%  & 17.1\% & 92.2\%  & 7.8\%  \\ \midrule
    
   \multicolumn{5}{c}{\textbf{Analysis by profession}} \\ \midrule
   \multirow{2}{*}{\textbf{Profession}} & \multicolumn{2}{c|}{\textbf{Reproducibility Status}} & \multicolumn{2}{c}{\textbf{Reproducibility Challenges}} \\ 
                                    & \textbf{Irreproducible}   & \textbf{Reproducible}   & \textbf{Agree}   & \textbf{Disagree} \\ \midrule
    SD      & 86.5\%  & 13.5\%  & 89.8\%  & 10.2\% \\ \midrule
    AP  & 98.8\%  & 1.2\%  & 85.6\%  & 14.4\% \\ \midrule
    TL \& RE & 69.5\%  & 30.6\%  & 75.8\%  &  24.2\% \\ \bottomrule
    \end{tabular}
    }
\end{table}

We then analyze the agreement based on the participants' Java experience and profession. Table \ref{table:agreement-experience-profession} summarizes the results. 
Participants with more experience (over five years) agree 8\% more on average with the reproducibility status than those with less experience (five years or less). 
Conversely, more experienced participants agree 7\% less with the reproducibility challenges. Experienced participants might have the skills to reproduce issues, while less experienced participants are more eager to resolve challenges (e.g., finding external libraries and fixing database dependencies).
Profession-wise analysis shows that technical leads and research engineers agree less with reproducibility status and challenges. They might not be actively programming and have a hectic task schedule, leading to guesses about the reproducibility status and challenges. They mainly disagree on the reproducibility status and challenge of question 4 (see Fig. \ref{fig:reproducibility-challenge-4}), where the issue was "outdated code." 

\begin{table}[htb]
	\centering
	\caption{Combined reproducibility challenges from Mondal et al. (C\textsubscript{1}-C\textsubscript{7}) and the survey (C\textsubscript{8}-C\textsubscript{10})}
	\label{table:combined-reproducibility-challenges}
    \resizebox{3.7in}{!}{%
    \begin{tabular}{p{11cm}}  \toprule

    (C\textsubscript{1}) Class/Interface/Method not found \\ 
    (C\textsubscript{2}) Important part of code missing \\
    (C\textsubscript{3}) External library not found \\
    (C\textsubscript{4}) Identifier/Object type not found \\
    (C\textsubscript{5}) Too short code snippet \\
    (C\textsubscript{6}) Database/File/UI dependency \\
    (C\textsubscript{7}) Outdated code  \\ 
    (C\textsubscript{8}) Error log/stack trace missing  \\
    (C\textsubscript{9}) System dependency or environment setup is missing   \\
    (C\textsubscript{10}) Sample input-output missing \\ \bottomrule
    
    \end{tabular}
    }
\end{table}

In addition to the given challenges, participants report three additional challenges they encountered.

    (i) \emph{Error log/stack trace missing.} Error logs or stack traces provide meaningful insights into program failures. Some issues cannot be reproduced without them.
    
    (ii) \emph{System dependency or environment setup is missing.} Some programming issues are tied to specific Operating Systems (e.g., Windows), IDEs (e.g., Eclipse), and software versions (e.g., Python 3). These issues cannot be reproduced when questions miss those details.
    
    (iii) \emph{Sample input-output missing.} Several issues require sample input-output to be reproduced.

\noindent Table \ref{table:combined-reproducibility-challenges} presents all reproducibility challenges, combining those from Mondal et al. with additional challenges from the survey.

\smallskip
\noindent\textbf{Summary.} On average, about 88\% of participants cannot reproduce the issues using the code snippets, and 89\% encounter the given challenges. Only about 11\% claim they could reproduce the issues and thus disagree with the challenges. 
Our analysis by experience and profession shows that the lowest agreement level is 70\%, which is acceptable. Such agreements support and validate the reproducibility status and challenge catalog (Table \ref{table:reproducibility-challenges}).

\begin{figure}[!htb]
\centering
   	\pgfplotstableread{
    		1	7.5   5.6   13.2  3.8   11.3  9.4   20.8
    		2	35.9  15.1  26.4  34    7.5   28.3  22.6
    		3	24.5  17    32.1  43.4  30.2  34    22.6
    		4	30.2  58.5  24.5  9.4   47.2  22.6  17
    		5	1.9   3.8   3.8   9.4   3.8   5.7   17

    	}\datatable
    	
      \resizebox{5.8in}{!}{%
      \begin{tikzpicture}
        	\begin{axis}[
        	xtick=data,
        	xticklabels={No Problem, Moderate, Severe, Blocker, No Opinion
        	},
        	enlarge y limits=false,
        	enlarge x limits=0.12,
        	ymin=0,ymax=100,
        	ybar,
        	bar width=0.33cm,
        	width=6.7in,
        	height = 3.6in,
        	ytick={0,20,...,100},
            yticklabels={0\%,20\%,40\%,60\%,80\%,100\%},
        	ymajorgrids=false,	
        	major x tick style = {opacity=0},
        	minor x tick num = 1,    
        	minor tick length=1ex,
        	legend style={
                draw=none,
                fill=none,
        	at={(0.45, 0.98)},
        	font=\small,
        	legend cell align=left,
            anchor=north,
            legend columns=2
            },
            legend image code/.code={
                \draw (0cm,-0.1cm) rectangle (1em,0.4em);
            },
            nodes near coords style={rotate=90,  anchor=west, font=\small},
        	nodes near coords =\pgfmathprintnumber{\pgfplotspointmeta}\%
        	]
        	\addplot[draw=black!80, fill=black!0] table[x index=0,y index=1] \datatable;
        	\addplot[draw=black!80, fill=black!2, postaction={pattern=horizontal lines}] table[x index=0,y index=2] \datatable;
        	\addplot[draw=black!80, fill=black!4, postaction={pattern=vertical lines}] table[x index=0,y index=3] \datatable;
        	\addplot[draw=black!80, fill=black!6, postaction={pattern=grid}] table[x index=0,y index=4] \datatable;
        	\addplot[draw=black!80, fill=black!8, postaction={pattern=dots}] table[x index=0,y index=5] \datatable;
        	\addplot[draw=black!80, fill=black!10, postaction={pattern=north east lines}] table[x index=0,y index=6] \datatable;
        	\addplot[draw=black!80, fill=black!12, postaction={pattern=north west lines}] table[x index=0,y index=7] \datatable;
            \legend	{Class/Interface/Method not found,
            		 Important part of code missing,
            		 External library not found,
            		 Identifier/Object type not found,
            		 Too short code snippet,
            		 Database/File/UI dependency,
            		 Outdated code
            		 }
        	\end{axis}
    	\end{tikzpicture}
    	}
\caption{Impacts on the reproducibility challenges answering questions.}
\label{fig:impacts-reproducibility-challenges}
\end{figure}
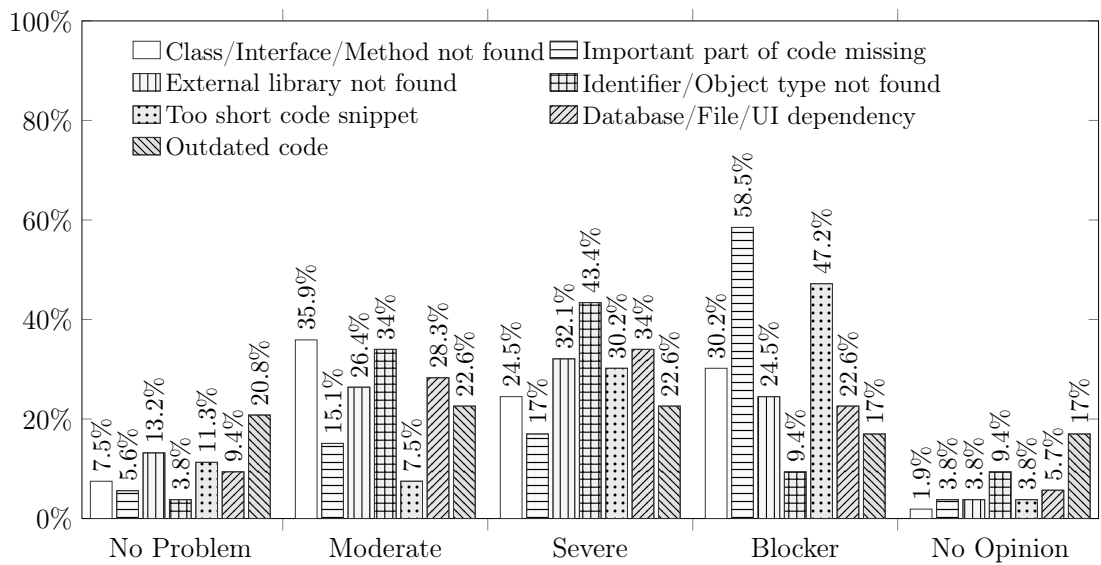

\subsection{Impacts of Each of the Reproducibility Challenges to Answer Questions (RQ2)}
\label{impact-analysis}

We analyze practitioners' agreement with the reproducibility challenge in Section \ref{agree-analysis-rq1}. However, understanding how these challenges affect answering questions is crucial for determining which ones need to be addressed early.
Fig. \ref{fig:impacts-reproducibility-challenges} shows how participants assess the impact of each reproducibility challenge. The majority (about 36\%) perceive ``Class/Interface/Method not found'' as moderate, with around 30\% considering it a blocker. About 59\% of participants consider ``An important part of code missing'' a blocker. ``Too short code snippet'' is also considered a blocker by around 47\% of participants. Regarding ``Identifier/Object type not found,'' approximately 43\% perceive it as severe, while 34\% perceive it as moderate.

\begin{table*}[!htb]
\centering
	\caption{Practitioners' comments (unaltered) behind the choices of impacts to answer questions}
	\label{table:developers-comments-severity}
 	\resizebox{5.7in}{!}{%
    \begin{tabular}{p{16cm}} \toprule
    \emph{``I really think not finding `Class/Interface/Method' is a blocker problem, because how can I understand a code without their method explained and reproduce. Same case happens in the case of  `Too short code snippet' and 'Important part of code missing'.''} \\ \midrule 
    
    \emph{``Without Class/Method definition, it's nearly impossible to identify the exact error. Besides, it is helpful, if individuals give some sample input and output examples. Dependency on Database: this is kind of blocker to run the sample code. ''} \\ \midrule 

    \emph{``If the given code part is too small, solution can not be made by depending only few line of code.''} \\ \midrule 
    
    \emph{``As I already answered most of the time we are irritated to help with question without proper playground. ( class not found, Important code missing... etc.).''} \\ \bottomrule 
\end{tabular}
}
\end{table*}

We then attempt to identify the top challenges in each of the five categories. According to Fig. \ref{fig:impacts-reproducibility-challenges}, ``Class/Interface/Method not found'' is the top moderate challenge category. In the severe category, ``Identifier/Object type not found'' ranks highest, while ``Important part of code missing'' ranks highest in the blocker category. 
While ``Outdated code'' received the highest percentage in the `no problem' and `no opinion' categories, participants also perceived this challenge as a moderate (22.6\%) or severe (22.6\%) problem.
Migrating outdated code or utilizing alternative classes/APIs could mitigate this challenge, reducing its impact on developers.
%
We also attempt to gain more insights by asking practitioners for justification for their choices. Even though it was optional, we received a few of their comments (see Table \ref{table:developers-comments-severity}). For example, one developer noted that without the Class/Method definition, reproducing the exact issues is nearly impossible.

\smallskip
\noindent\textbf{Summary.} Most developers could not reproduce issues and thus could not answer questions if important code statements were missing. Dealing with other challenges, such as Class/Interface/Method and Identifier/Object type not found, could irritate and waste developers' valuable time and prevent/delay appropriate solutions. However, dealing with outdated code is relatively easy.

\begin{figure*}[!htb]
\centering
   	\pgfplotstableread{
    		1	72.4  95.2   83.8  84.6   100  75   
    		2	10.3  14.3   22.2  0.0    27.3 0    

    	}\datatable
    	
      \resizebox{4in}{!}{%
      \begin{tikzpicture}
        	\begin{axis}[
        	xtick=data,
        	xticklabels={Unresolved, Unanswered
        	},
        	xticklabel style={font=\large},
        	enlarge y limits=false,
        	enlarge x limits=0.40,
        	ymin=0,ymax=125,
        	ybar,
        	bar width=0.65cm,
        	width=5.5in,
        	height = 3.3in,
        	ytick={0,20,...,100},
            yticklabels={0\%,20\%,40\%,60\%,80\%,100\%},
        	ymajorgrids=false,
        	yticklabel style={font=\large},	
        	major x tick style = {opacity=0},
        	minor x tick num = 1,    
        	minor tick length=1ex,
        	legend style={
                draw=none,
                fill=none,
        	at={(0.70, 0.95)},
        	font=\small,
                legend cell align=left,
                anchor=north,
            },
           legend image code/.code={
                \draw (0cm,-0.1cm) rectangle (1em,0.4em);
            },
            nodes near coords style={rotate=90,  anchor=west, font=\small},
        	nodes near coords =\pgfmathprintnumber{\pgfplotspointmeta}\%
        	]
        	\addplot[draw=black!80, fill=black!0] table[x index=0,y index=1] \datatable;
        	\addplot[draw=black!80, fill=black!2, postaction={pattern=horizontal lines}] table[x index=0,y index=2] \datatable;
        	\addplot[draw=black!80, fill=black!4, postaction={pattern=vertical lines}] table[x index=0,y index=3] \datatable;
        	\addplot[draw=black!80, fill=black!6, postaction={pattern=grid}] table[x index=0,y index=4] \datatable;
        	\addplot[draw=black!80, fill=black!8, postaction={pattern=dots}] table[x index=0,y index=5] \datatable;
        	\addplot[draw=black!80, fill=black!10, postaction={pattern=north east lines}] table[x index=0,y index=6] \datatable;
            \legend	{Class/Interface/Method not found,
            		 Important part of code missing,
            		 External library not found,
            		 Identifier/Object type not found,
            		 Too short code snippet,
            		 Database/File/UI dependency,
            		 Outdated code
            		 }
        	\end{axis}
    	\end{tikzpicture}
    	}
\caption{Percentage of unresolved \& unanswered questions vs reproducibility challenges.}
\label{fig:severity-reproducibility-challenges-answer-metadata}
\end{figure*}
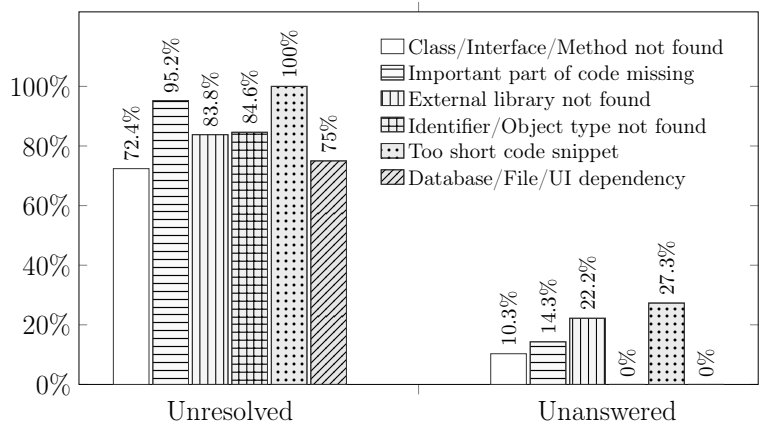

\begin{figure}[htb]
	\centering
	\resizebox{5.5in}{!}{
	\subfloat[]
    {
	\resizebox{4in}{!}{
    \includegraphics[width=3in]{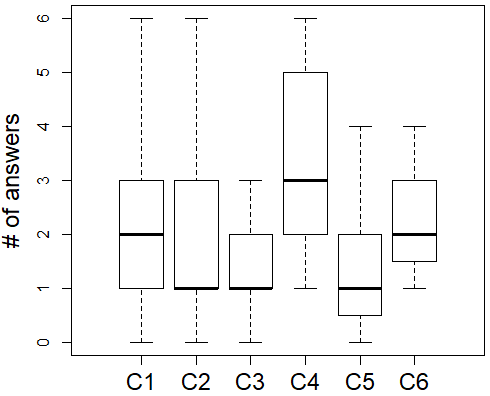}
    \label{fig:no-of-answers}
    }
    }
    \subfloat[]
    {
	\resizebox{4in}{!}{
    \includegraphics[width=3in]{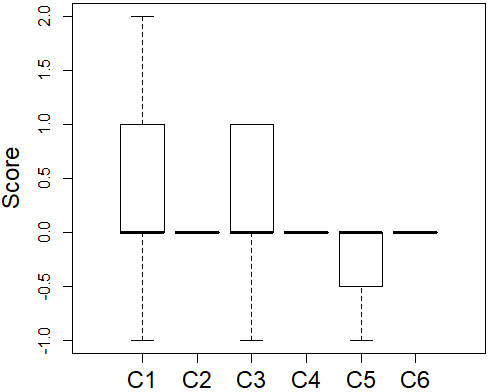}
    \label{fig:score}
    }
    }
    }
\caption{Answers \& score vs reproducibility challenges (\textbf{C1:} Class/Interface/Method not found, \textbf{C2:} Important part of code missing, \textbf{C3:} External library not found, \textbf{C4:} Identifier/Object type not found, \textbf{C5:} Too short code snippet, \textbf{C6:} Database/File/UI dependency)}

\label{fig:answer-score-vs-challenge}
\end{figure}

\subsection{Agreement Between Practitioners' Opinion \& Empirical Findings on Impact Analysis (RQ3)}

In the previous section, we saw practitioners' opinions on the impact of each reproducibility challenge. This section examines how much empirical evidence supports these opinions.

The dataset published by Mondal et al.~\citep{Mondal-SOIssueReproducability-MSR2019} contains 87 questions with irreproducible issues. We first categorize these questions based on their challenges. Then, we collect the question meta-data (e.g., presence of accepted answers, no of answers, and score) and compare this meta-data against question categories to see the impact of these challenges. We skip the question with \emph{outdated code} challenge due to its single occurrence in the dataset.

Fig. \ref{fig:severity-reproducibility-challenges-answer-metadata} shows the percentage of unresolved (questions without acceptable answers) and unanswered questions against each challenge type.
All of the questions associated with \emph{too short code snippet} challenge remained unresolved. For \emph{important part of code missing}, 95.2\% of questions remained unresolved. These findings align with RQ2 where \emph{important part of code missing} \& \emph{too short code snippet} challenges were considered blockers. 
Additionally, 27.3\% of questions were unanswered due to \emph{too short code snippet}, with the second highest unanswered questions attributed to \emph{external library not found} challenge. Interestingly, \emph{External library not found} was also assessed as the second most severe challenge in RQ2.

Fig. \ref{fig:no-of-answers} shows the box plots representing the number of answers against each challenge. The median number of answers is only one for the reproducibility challenges \emph{important part of code missing}, \emph{too short code snippet}, and \emph{external library not found}. These findings suggest that these challenges significantly affect the chance of getting answers, supporting the practitioners' opinion in RQ2.

We then analyze the quality of questions against each reproducibility challenge. In SO, the quality of a post is subjectively approximated by a metric called \emph{score} (i.e., upvotes $-$  downvotes) \citep{mondal2021rollback, yao2015detecting,neshati2017early,duijn2015quality}. Fig. \ref{fig:score} shows the box plots of question scores vs. reproducibility challenges. The median score for questions is consistently zero, regardless of the reproducibility challenges. This finding suggests that questions with irreproducible issues struggle to attract user interest for upvotes, as they are often perceived as low quality.

\smallskip
\noindent\textbf{Summary.} Empirical evidence confirms the practitioners' opinion on the impacts of reproducibility challenges on answering questions. The analysis between challenge-based question category vs. question meta-data shows that \emph{important part of code missing}, \emph{too short code snippet}, and \emph{external library not found} significantly hinder questions from receiving acceptable answers or any answers at all.

\subsection{Severity Analysis of the Reproducibility Challenges (RQ4)}

It is not always possible to fix all the reproducibility challenges. Thus, it is important to prioritize more severe challenges and fix them within a strict budget (e.g., limited time).



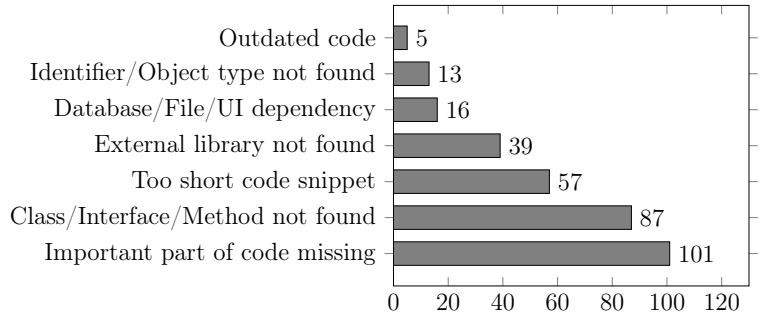
\begin{figure}[!htb]
\centering
        \resizebox{4in}{!}{%
    	\begin{tikzpicture}
        	\begin{axis}
            [
            height = 2.5in,
            width = 3in,
            xmin=0, xmax=130,
            xbar,
            ytick = data,
            enlarge y limits=0.15,
        	enlarge x limits=false,
	        nodes near coords,
            bar width=0.4cm,
            symbolic y coords = {
            Important part of code missing,
            Class/Interface/Method not found,
            Too short code snippet,
            External library not found,
            Database/File/UI dependency,
            Identifier/Object type not found,
            Outdated code
            },
            legend style={at={(0.5,-0.20)},
        	font=\footnotesize,
            anchor=north,legend columns=-1}
            ]
            \addplot[xbar,fill=black!50] coordinates {
            (101,Important part of code missing)
            (87,Class/Interface/Method not found)
            (57,Too short code snippet)
            (39,External library not found) 
            (16,Database/File/UI dependency) 
            (13,Identifier/Object type not found) 
            (5,Outdated code) 
            };
           \end{axis}
        \end{tikzpicture}
        }
\caption{The rank of reproducibility challenges according to their score.}
\label{fig:severity-reproducibility-challenges}
\end{figure}

We employ the Borda count~\citep{Yamashita-DoDevelopersCareAboutCodeSmell-WCRE2013} to analyze and rank the challenges based on their severity. This method assigns points to candidates based on their rank order: the first-ranked candidate receives the highest points, decreasing sequentially. In our case, with three options to choose from, the first option receives 3 points, the second 2 points, and the third 1 point. 
Fig. \ref{fig:severity-reproducibility-challenges} shows the severity points for each challenge. The ``important part of code missing'' receives the highest (101), while ``Outdated code'' gets the lowest (5) point. Based on our analysis, the top three priorities are -- (1) important part of code missing, (2) class/interface/method not found, and (3) too short code snippet. These findings match those from the impact analysis (see Section \ref{impact-analysis}).

\smallskip
\noindent\textbf{Summary.} Based on the severity analysis, question submitters should include crucial code parts that cannot be guessed and define necessary classes/methods. They should also avoid submitting code snippets that are too short (e.g., 1-2 lines) to get quick and appropriate solutions.

\begin{table*}[htb]
\centering
	\caption{Practitioners' tool support recommendations (unaltered) to promote reproducibility}
	\label{table:tool-support-recommendations}
 	\resizebox{5.7in}{!}{%
    \begin{tabular}{p{16cm}} \toprule
    \emph{``The tool that helps me to guess the missing code, library version, missing input data depending on the issue context. It will be great if the tool generates missing information for me depending on the question/issue.''} \\ \midrule 
    
    \emph{``A software that can ensure the missing cases and force the user to complete them, Auto code generator.''} \\ \midrule 

    \emph{``A tool that helps to guess the missing important code depending on the question context.''} \\ \midrule 
    
    \emph{``A tool that can sense my written code snippet and see what important part  might be missing there.''} \\ \midrule 
    
    \emph{``It will provide suggestions to reproduce after pasting the code in IDE. ''} \\ \bottomrule 

\end{tabular}
}
\end{table*}

    

\begin{table}[htb]
\centering
	\caption{Assessment of the tool support options}
	\label{table:tool-support-options}
 	\resizebox{5in}{!}{%
    \begin{tabular}{p{12cm}|c} \toprule
    
    \multicolumn{1}{c|}{\textbf{Options}} & \multicolumn{1}{c}{\textbf{Mean Value}} \\ \midrule
    (1) A tool that suggests users improve the code examples to support reproducibility                             &   4.43     \\ \midrule 
    (2) A tool that warns users about reproducibility challenges that are severe or may block the reproducibility   &   4.25     \\ \midrule 
    (3) An IDE (e.g., Eclipse) plugin to support reproducibility                                                    &   4.25     \\ \midrule 
    (4) A browser plugin to support reproducibility                                                                 &   4.02     \\ \midrule 
    (5) A website that guides users to improve reproducibility                                                      &   3.85     \\ \midrule 
    (6) A static analyzer that examines code snippets to find reproducibility challenges                            &   4.15    \\ \bottomrule 
\end{tabular}
}
\end{table}

\subsection{Tool Support and Design Requirements to Promote Reproducibility (RQ5)}

To improve question quality by assuring reproducibility for quicker, appropriate answers, we plan to introduce interactive tools (e.g., browser/IDE plugins) and seek practitioners' opinions.
Table \ref{table:tool-support-recommendations} presents a few valuable recommendations from participants (the complete list is available in our online appendix \citep{ourdataset}). For example, one participant suggested ``A tool that can sense my written code snippet and see what important part might be missing there''.

We received 29 responses (out of 53) with practical tool design recommendations. We disregarded the rest as they were either too generic or unclear. The first author conducted an opening coding to summarize these recommendations as follows.

\smallskip
$\bullet$ \emph{Plugin/Linter/Live code running environment.}
About 52\% (14 out of 29) of participants suggested adding a browser/IDE plugin, linter, or live coding environment. They specifically requested tools to check the parsability/executability of the code snippets, crucial for reproducing question issues \citep{mondal2022reproducibility}. These tools can help improve code snippets for better reproducibility.

\smallskip
$\bullet$ \emph{Missing code finder.}
About 24\% of participants propose implementing a tool that engages with question submitters when code snippets miss an important part that is mandatory to reproduce the issue.

\smallskip
$\bullet$ \emph{Notify if the library inclusion is missed.}
Finding external libraries is a major challenge, especially when they are not imported in code snippets. About 21\% of survey participants recommend tools that help them include these libraries as needed, which can save time.

\smallskip
$\bullet$ \emph{Environmental setup-related information finder.}
About 7\% of the participants asked for a tool to analyze the questions' texts to identify missing environmental setup details (e.g., software version, Operating System, IDE), which are crucial for reproducibility.

\smallskip
$\bullet$ \emph{Missing input-output detector.}
According to the survey, 6.9\% of participants suggest tools for identifying necessary input-output examples to reproduce issues.

\smallskip
$\bullet$ \emph{Auto code generator.}
About 7\% of participants suggested an auto code generator to generate missing code or search for similar code snippets for reproducibility.

$\bullet$ \emph{Missing definition/declaration finder.}
3.4\% of participants suggested a tool to add missing class/method definitions and identifier/object declarations in code snippets, which could promote reproducibility.

$\bullet$ \emph{Remind if the included code snippet is too short.}
From the responses, 3.4\% of participants want tools to warn question submitters if they include code snippets that are too short. It is hard to identify the problem accurately from such brief code, and guesses often lead to incorrect answers that do not solve the actual problem.

\smallskip
$\bullet$ \emph{Code readability estimator.}
A few participants (3.4\%) suggest tools to estimate the readability of the code snippets. Poor readability hinders code comprehension and reproducibility.

\smallskip
Additionally, Table \ref{table:tool-support-options} lists our six tool options. Participants rated the first three options as ``highly influential'' (mean value $\geqslant$ 4.21) and the remaining three as ``influential'' (3.41 $\leqslant$ mean value $\leqslant$ 4.20).

\smallskip
\noindent\textbf{Summary.} Practitioners prefer browser/IDE plugins that suggest improving code examples for better reproducibility, such as identifying crucial missing code and external libraries. They also recommend tools that assess reproducibility challenges by severity.

\section{Prediction of Reproducibility Status}
\label{sec:reproducibility-estimation}

To introduce recommended tool supports, we first attempt to develop ML models to predict whether issues are reproducible. The steps for developing, evaluating, interpreting, and validating ML models are discussed below.

\subsection{Selection of Dataset}
\label{subsec:dataset-selection}

We use Mondal et al.'s dataset \citep{Mondal-SOIssueReproducability-MSR2019}, which consists of 357 questions (270 reproducible + 87 irreproducible), to extract features and develop ML models.

\subsection{Feature Extraction}
\label{subsec:feature-extraction}

We extract nine source code-related features to predict the reproducibility of issues using ML models. We focus on the \emph{executability} and \emph{challenges that prevent reproducibility} to extract these features. The nine features are discussed below.

\begin{figure}[!htb]
      \centering
      \includegraphics[width=2.5in]{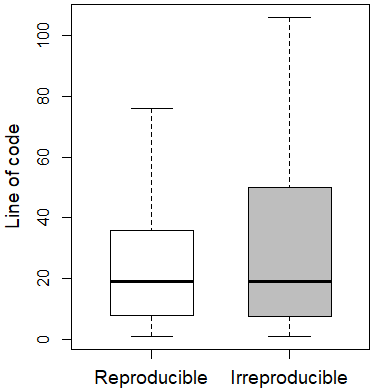}
      \caption{Line of code of the code snippets included in the questions with reproducible \& irreproducible issues.}
      \label{fig:loc}
    \end{figure}

\textbf{Line of Code (LOC):} We extract code snippets from questions using \texttt{<code>} tags under \texttt{<pre>}, remove blank lines, and count the lines of code (LOC) to assess their length. Fig. \ref{fig:loc} shows that while the median LOC is similar (e.g., $LOC = 19$) for both reproducible and irreproducible issues, the average LOC is higher for irreproducible issues (i.e., $Avg (LOC) = 53.28$) compared to reproducible ones (i.e., $Avg (LOC) = 32.43$). Notably, 13.8\% (12 out of 87) of code snippets in questions with irreproducible issues are very short (1--2 LOC), whereas this figure is 5.9\% (16 out of 270) for questions with reproducible issues. Thus, LOC can be an important estimator of a code snippet's ability to reproduce reported issues.

\textbf{Presence of a Method:} We check if code snippets include user-defined methods. A method is a block of code designed for specific tasks. Including methods helps reproduce an issue when it is related to that task. In our dataset, about 67\% of code snippets from questions with reproducible issues include methods, compared to 55\% from questions with irreproducible issues.

\textbf{Presence of Main Method:} The main method can help code snippets be executable and reproducible issues. In our dataset, 55\% of code snippets from questions with reproducible issues include the main method, compared to 21\% for questions with irreproducible issues.

\textbf{Presence of a Class:} Many Java code snippets are just a few bare statements without being encapsulated correctly in a class, which is essential for parsing or compiling. Yang et al.~\citep{querytousablecode} improved parse rates by 13.02\% and compilation rates by 1.42\% by encapsulating snippets with a dummy class when absent. Missing this class is a key challenge affecting reproducibility \citep{emse2021bmondal}. In our dataset, 54\% of code snippets reproducing reported issues included a class, compared to 39\% for code snippets with irreproducible issues.

\textbf{Parsability:} We examine how well code snippets can be parsed. Using \texttt{Java Development Kit (JDK) 1.8} and \texttt{JavaParser}\footnote{\url{http://javaparser.org}}, we can successfully parse 50.7\% of code snippets that reproduce reported issues, and 37.9\% of code snippets that do not.

\textbf{Compilability:} Successfully executing code snippets is crucial for reproducing reported issues unless the issue is related to a compilation error. We use \texttt{javac} to check if code snippets compile successfully. Our analysis shows that 31.1\% of code snippets with reproducible issues can be compiled successfully, compared to 1.1\% with irreproducible issues.

\textbf{Inclusion of Native Library:} It is crucial to include necessary libraries for compiling and executing code to reproduce issues. IDEs (e.g., Eclipse) often suggest native libraries for common tasks. However, many users copy code into a text editor and then try to execute it from the command line to reproduce issues. Thus, they recommend adding import statements for these libraries. Our feature values are defined as follows.

(a) $-1$: Required import statements for native libraries were absent.

(b) $0$: No native library was needed, or it was unclear from the code snippets.

(c) $1$: Necessary import statements for native libraries were present.

\noindent Our analysis finds that 51\% (82 out of 162) of code snippets with reproducible issues include the required import statements, while only 34\% (16 out of 46) of code snippets with irreproducible issues do so.

\textbf{Inclusion of External Library:} We follow an approach similar to \emph{Inclusion of Native Library} to determine feature values for \emph{Inclusion of External Library} as follows. 

(a) $-1$: Required import statements for external libraries were absent.

(b) $0$: No external library was needed, or it was unclear from the code snippets.

(c) $1$: Necessary import statements for external libraries were present.


\textbf{Exception Handling:} In some cases, unhandled exceptions could prevent issue reproducibility unless the issue is related to those exceptions. We thus check if the code throws any unexpected exceptions and assign feature values as follows:

(a) $-1$: Exceptions were not handled, but we could handle them.

(b) $0$: No exception handling was needed to compile/execute the code snippets, or it was unclear from the code snippets.

(c) $1$: Necessary exceptions were handled correctly, or the reported issues were related to unexpected exceptions.

\subsection{Resolve Class Imbalance Problem}
\label{subsec:class-imbalance-problem}

Our dataset (Section \ref{subsec:feature-extraction}) has a class imbalance problem, with 270 reproducible samples and only 87 irreproducible samples. Standard ML techniques (e.g., Random Forest) are biased towards the majority class and often misclassify the minority class. To address this, we use the Synthetic Minority Oversampling Technique (SMOTE) \citep{chawla2002smote}, which creates synthetic examples for the minority class. We oversample the \emph{irreproducible} class to equalize it with the \emph{reproducible} class.

\subsection{Machine Learning Model Selection}
\label{subsec:model-selection}

Studies suggest that supervised ML classifiers (e.g., Random Forest) can perform equally well or even better than deep learning (DL) techniques \citep{fu2017easy, beyer2018automatically} for small datasets. Moreover, DL techniques are more complex, costly, and prone to overfitting with small datasets \citep{beyer2018automatically}. Therefore, we chose five popular supervised ML classifiers with different learning strategies to classify the issue's reproducible/irreproducible status. These algorithms are selected because (1) they can build reliable models to predict reproducibility status using our features, and (2) they are widely used in relevant studies \citep{mondal2023automatic, TUT-Saha-2013, UAC-Ponzanelli-2014, AII-Rahman-2015, beyer2018automatically}.

\smallskip\noindent\textbf{Random Forest (RF)} is a popular supervised ML technique due to its simplicity and versatility \citep{randomForest}. It creates an ensemble of decision trees, namely `forest', trained with the `bagging' method. The idea behind ensemble learners is that multiple weak learners combine to form a strong learner to improve performance \citep{EBS-Polikar-2006}. RF scales well to any number of dimensions and offers acceptable performance. Instead of searching for the most important feature when splitting nodes, it looks for the best feature within a random subset \citep{mondal2023automatic}, which helps prevent model overfitting.

\smallskip\noindent\textbf{eXtreme Gradient Boosting (XGBoost)} is a scalable tree-boosting technique \citep{chen2015xgboost} that predicts a target variable by combining estimates from simpler weak models. It works for classification and regression and offers improved speed and performance compared to traditional gradient-boosted decision trees (GBM) \citep{mondal2023automatic}. XGBoost supports parallel and distributed computing, making it faster than other algorithms, and handles different data types and relationships well. It also includes built-in cross-validation and various regularization techniques to prevent overfitting.

\smallskip\noindent\textbf{Artificial Neural Network (ANN)} mimics the brain's neural networks. Unlike perceptrons, which only handle linearly separable data, ANN (Multi-layered Perceptrons) can manage non-linearly separable data. It consists of interconnected neurons arranged in layers. ANN depends on three fundamental aspects: (1) input and activation functions, (2) network architecture, and (3) weights of each input connection.
The behavior of an ANN depends on its current connection weights, which start as random values. During training, instances from the training set are repeatedly exposed to the network. The network adjusts its weights slightly to minimize the difference between its output and the desired values.

\smallskip\noindent\textbf{Naive Bayes (NB)} is a simple yet popular and effective ML classifier. It assumes that input features are independent. It is a probabilistic algorithm based on Bayes' Theorem, which uses the Maximum A Posteriori decision rule in Bayesian settings for classification. NB can also be represented as a basic Bayesian network.

\smallskip\noindent\textbf{K-Nearest Neighbors(KNN)} is a non-parametric technique for classification and regression problems \citep{NCA-Goldberger-2005}. It does not use the training dataset for generalization, making the training phase faster than other algorithms. In KNN, K determines how many nearby neighbors influence the classification of a data point. It finds the closest K neighbors to a target point using measures like Euclidean distance. Each neighbor votes for its class, and the class with the most votes predicts the outcome.

\subsection{Performance Analysis}
\label{subsec:performance-analysis}

We use default settings from \emph{Scikit-learn} \citep{SML-Pedregosa-2011} for our models and apply 10-fold cross-validation. This method splits the data into ten folds, training on nine and testing on one fold iteratively until all the data are used for training and testing. Cross-validation is one of the most widely used data resampling methods to assess the model's generalization ability and to prevent overfitting \citep{hastie2009elements, hart2000pattern, berrar2019cross}. 
We then evaluate the performance of our prediction models using four performance metrics: (1) \emph{precision} that measures the ratio of correctly classified reproducibility status against all samples classified into that class, (2) \emph{recall} that measures the ratio of correctly classified reproducibility status against the observed samples as true instances, (3) \emph{F1-score} that offers the harmonic mean of precision and recall, and (4) \emph{classification accuracy} that measures the ratio of correctly classified reproducibility status into true \& false classes against all samples. These metrics give a clear picture of how well our models perform.

\begin{table}[htb]
\centering
\caption{Experimental Results}
\label{table:result-analysis}
\resizebox{5in}{!}{%
\begin{tabular}{l|l|c|c|c|c}
\toprule
\multicolumn{1}{c|}{\textbf{Technique}} & \multicolumn{1}{c|}{\textbf{Category}} & \textbf{Precision} & \textbf{Recall} & \textbf{F1-Score} & \textbf{Accuracy} \\ \midrule

\multirow{2}{*}{\textbf{RF}} &  \textbf{Reproducible}     & 82.7\%  & \textbf{83.0\%}  & \textbf{82.8\%}  & \multirow{2}{*}{\textbf{82.8\%}}     \\ \cmidrule{2-5}
                             & \textbf{Irreproducible}    & \textbf{82.9\%}  & 82.6\%  & \textbf{82.8\%}  &                         \\ \midrule
                             
\multirow{2}{*}{\textbf{XGBoost}} & \textbf{Reproducible}     & 81.8\%  & 81.5\%  & 81.6\%  & \multirow{2}{*}{81.7\%}     \\ \cmidrule{2-5}
                                  & \textbf{Irreproducible}   & 81.6\%  & 81.9\%  & 81.7\%  &                         \\ \midrule
                                  
\multirow{2}{*}{\textbf{ANN}} & \textbf{Reproducible}        & \textbf{84.5\%}  & 66.7\%  & 74.5\%  & \multirow{2}{*}{77.2\%}     \\ \cmidrule{2-5}
                                  & \textbf{Irreproducible}  & 72.5\%  & \textbf{87.8\%}  & 79.4\%  &                         \\ \midrule
                                  
\multirow{2}{*}{\textbf{NB}}    & \textbf{Reproducible}       & 83.2\%  & 60.7\%  & 70.0\%  & \multirow{2}{*}{74.1\%}     \\ \cmidrule{2-5}
                                  & \textbf{Irreproducible}  & 68.9\%  & \textbf{87.8\%}  & 77.2\%  &                         \\ \midrule
                                  
\multirow{2}{*}{\textbf{KNN}}    & \textbf{Reproducible}     & 75.3\%  & 70.0\%  & 72.6\%  & \multirow{2}{*}{73.5\%}     \\ \cmidrule{2-5}
                                  & \textbf{Irreproducible}  & 72.0\%  & 77.0\%  & 74.4\%  &                         \\ \bottomrule
                                  
                                  
                                  
\end{tabular}
}
\end{table}

\subsection{Model Performance to Classify Reproducibility Status}
\label{subsec:model-performance}

Table \ref{table:result-analysis} summarizes our experimental results. Our models classify reproducible issues with 75.3\% -- 84.5\% precision, 60.7\% -- 83.0\% recall, and 70.0\% -- 82.8\% F1-score. The ANN model achieved the highest precision (84.5\%), while the RF model achieved the highest recall (83.0\%) and the highest F1-score (82.8\%). For irreproducible issues, our models achieved 68.9\% -- 82.9\% precision, 77.0\% -- 87.8\% recall, and 74.4\% -- 82.8\% F1-score. Overall accuracy is above 73\%, with the highest being 82.8\%. All models showed consistent performance, indicating the strength of our selected features. However, the NB model had comparatively low recall for reproducible issues due to its assumption of feature independence, which is not realistic in our case. Overall, the RF model performed the best. 

\begin{figure}[htb]
	\centering
	\includegraphics[width=5.5in]{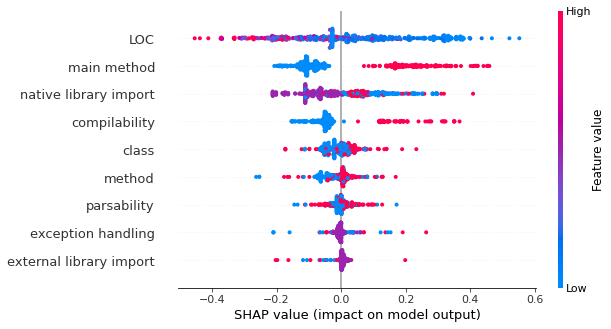}
	\caption{Feature importance using bee swarm plot (Random Forest model)}
	\label{fig:shap-summary-plot}
\end{figure}

\subsection{Model Performance Interpretation}
\label{subsec:model-performance-interpretation}

\begin{figure}[htb]
	\centering
	\includegraphics[width=5in]{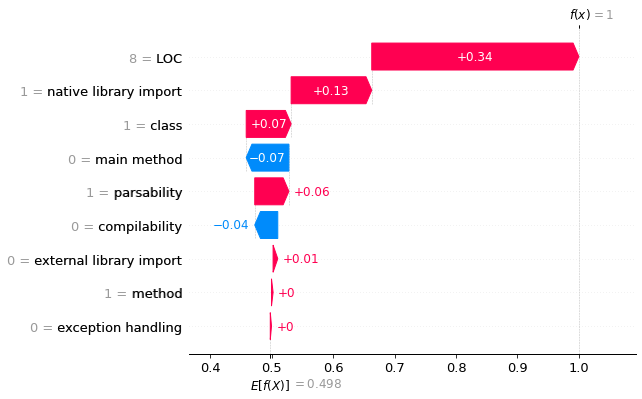}
	\caption{Features importance using waterfall plot (reproducible issue)}
	\label{fig:shap-waterfall-plot-reproducible}
\end{figure}

\begin{figure}[htb]
	\centering
	\includegraphics[width=5.7in]{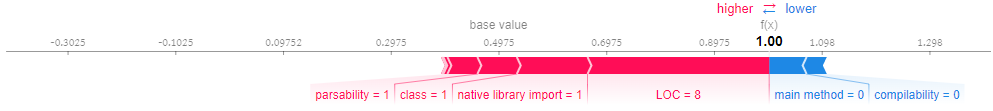}
	\caption{Feature importance using force plot (reproducible issue)}
	\label{fig:shap-force-plot-reproducible}
\end{figure}

\begin{figure}[htb]
	\centering
	\includegraphics[width=5in]{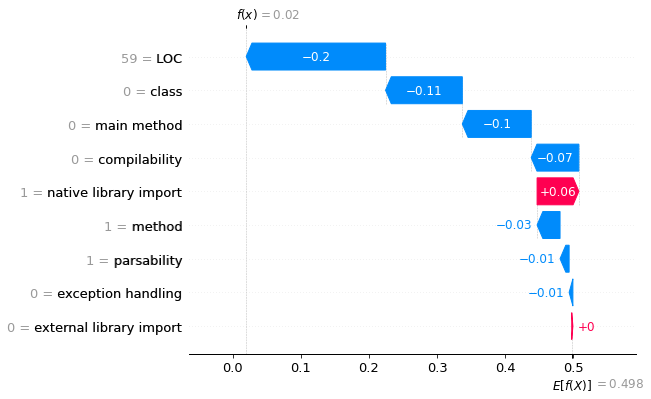}
	\caption{Features importance using waterfall plot (irreproducible issue)}
	\label{fig:shap-waterfall-plot-irreproducible}
\end{figure}

\begin{figure}[htb]
	\centering
	\includegraphics[width=5.7in]{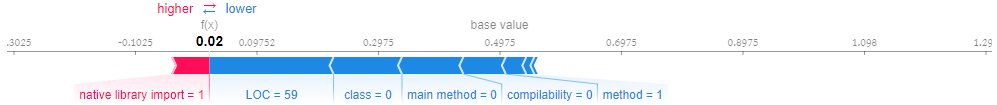}
	\caption{Feature importance using force plot (irreproducible issue)}
	\label{fig:shap-force-plot-irreproducible}
\end{figure}

Our classification models separate reproducible issues from irreproducible ones using nine features extracted from the provided code snippets in the questions. This section further explores how these features contributed to the classification. We use SHAP \citep{lundberg2017unified}, a popular model interpretation framework, to interpret the classification results of our models. The SHAP value is the average marginal contribution of a feature to the model's prediction across all combinations of features \citep{molnar2020interpretable, emse2022masud}. 
It shows whether a feature value increases a model's prediction over a random baseline \citep{lundberg2020local}. In our binary classification, \emph{issue reproducibility} is the \emph{positive} class, and \emph{issue irreproducibility} is the \emph{negative} class. Positive SHAP values indicate an increased prediction of the positive class and vice versa. Figures \ref{fig:shap-summary-plot}, \ref{fig:shap-waterfall-plot-reproducible}, \ref{fig:shap-force-plot-reproducible}, \ref{fig:shap-waterfall-plot-irreproducible} \& \ref{fig:shap-force-plot-irreproducible} summarize our analysis using SHAP values as follows.

Fig. \ref{fig:shap-summary-plot} shows the importance of our selected features using a bee swarm plot from our RF model. The bee swarm plot visualizes the SHAP value of a feature from each of the training instances on the x-axis. On the y-axis, it sorts all features in descending order based on their total SHAP values. In our dataset, true and false boolean responses are represented by 1 (true) and 0 (false). Blue indicates low feature values, while red indicates high feature values. Thus, small numerical values and false responses are represented in blue, and large numerical values and true responses are represented in red.

According to our RF model, the most important feature is LOC. That means the length of the code snippets in the questions is a key predictor of reproducibility. We observed that a high LOC often leads to negative SHAP values, indicating a higher likelihood of the issue being irreproducible. Our analysis shows that the average LOC for reproducible issues is 32, while it is 53 for irreproducible issues. However, a low LOC can also lead to negative SHAP values. We found that 13.8\% of irreproducible issues have code snippets that are too short (LOC = 1-2), compared to only 5.9\% for reproducible issues. This suggests that both too-short and too-long code snippets can hinder reproducibility. This observation can be explained in two ways. First, very short code snippets might lack enough detail to make them parsable/compilable, making it hard to reproduce the issue. Second, reproducing the issue from very long code snippets with multiple errors and external dependencies could be tedious.

We also analyze the following three most important features – \emph{main method}, \emph{native library import}, and \emph{compilability}. Their true responses lead to positive SHAP values, increasing the model's prediction toward issue reproducibility. In section \ref{subsec:feature-extraction}, we see that the main method appears more often in code snippets that reproduce issues. About 55\% of these snippets have the main method, compared to only 21\% for code snippets that do not reproduce issues. Similarly, code snippets with reproducible issues are 1.5 times more likely to include the required import statements for native libraries. Additionally, 31.1\% of snippets with reproducible issues can be compiled successfully, whereas only 1.1\% of snippets with irreproducible issues can be compiled. Thus, the presence of the main method, import statements for libraries, or successful compilation might indicate issue reproducibility. Similarly, \emph{presence of method \& class}, and \emph{parsability} with their true responses will likely increase our model's prediction towards issue reproducibility.

To verify these findings, we analyze one reproducible issue\footnote{\url{https://stackoverflow.com/questions/975619}} and one irreproducible issue\footnote{\url{https://stackoverflow.com/questions/790321}}. The waterfall (Fig. \ref{fig:shap-waterfall-plot-reproducible}) and force plots (Fig. \ref{fig:shap-force-plot-reproducible}) show how different features contribute to classifying an issue as reproducible. In Fig. \ref{fig:shap-waterfall-plot-reproducible}, we see that LOC ($8$), the presence of import statements for native libraries (\emph{native library import} $= 1$), the presence of a class (\emph{class} $= 1$), and parsability success (\emph{parsability} $= 1$) push the model’s prediction towards reproducibility. The presence of a method (\emph{method} $= 1$) also helps. However, the absence of the main method and compilation failure decrease the model’s prediction of reproducibility, as shown by their negative SHAP values.

The force plot in Fig. \ref{fig:shap-waterfall-plot-reproducible} also illustrates how several features (e.g., LOC, native library import, presence of class, and parsability success) increase our model’s prediction towards issues reproducibility, while the absence of the main method and compilation failure decrease it. These findings match our earlier observations (Fig. \ref{fig:shap-summary-plot}).
In Fig. \ref{fig:shap-waterfall-plot-irreproducible}, long code (LOC $= 59$), the absence of the main method and class, and compilation failure prevent the second code snippet\footnotemark[3] from being classified as reproducible. Fig. \ref{fig:shap-force-plot-irreproducible} supports this, though the presence of import statements for native libraries still has a positive SHAP value.


\begin{table}[htb]
\centering
	\caption{Impact of model features on issue reproducibility}
	\label{table:feature-impact}
 	\resizebox{3.5in}{!}{%
    \begin{tabular}{l|c|c|c} \toprule
    
    \textbf{Model Feature} & \textbf{$\chi^2$} & \textbf{DF} & \textbf{p-value} \\ \midrule
    
      \emph{LOC} & 6.4 & 2 & 0.04 $<$ 0.05  \\ \midrule 
      \emph{Parsability} & 3.8 & 1 & 0.05 $\leq$ 0.05 \\ \midrule 
      \emph{Compilability} & 30.9 & 1 & 2.67e-08 $<$ 0.05  \\ \midrule 
      \emph{Presence of method} & 3.3 & 1 & 0.07 $>$ 0.05   \\ \midrule 
      \emph{Presence of main method} & 28.8 & 1 & 7.95e-08 $<$ 0.05  \\ \midrule 
      \emph{Presence of class} & 4.8 & 1 & 0.03 $<$ 0.05  \\ \midrule 
      \emph{Native library import} & 4.7 & 2 & 0.09 $>$ 0.05  \\ \midrule 
      \emph{External library import} & 15.3 & 2 & 4.65e-4 $<$ 0.05  \\ \midrule 
      \emph{Exception handling} & 0.7 & 2 & 0.71 $>$ 0.05  \\ \bottomrule 

\end{tabular}
}
\end{table}

\subsection{Statistical analysis of the model features}
\label{suvsec:statistical-analysis}

We use five ML models to classify reproducible from irreproducible issues. We then identify the important features using their SHAP values. This section performs statistical analysis using significance tests to gain further insights into how these features affect issue reproducibility. We have eight categorical features and one numerical feature ($LOC$). We categorize LOC into three categories: short (below 25\textsuperscript{th} percentile), medium (25\textsuperscript{th} -- 75\textsuperscript{th} percentiles), and long (above 75\textsuperscript{th} percentile).  We use $\chi^2$-test \citep{mchugh2013chi} test to analyze the impact of features on issue reproducibility. 

Table \ref{table:feature-impact} summarizes our statistical findings. Six of our features (e.g., LOC, comparability) strongly correlate with issue reproducibility, as indicated by SHAP analysis. For example, the ability to compile code significantly affects issue reproducibility ($\chi^2$ value with 1 degree of freedom is 30.9, and p-value $\equiv 0 < 0.05$).
While SHAP analysis highlighted the importance of \emph{native library import} for model performance, our statistical tests found it insignificantly impacts issue reproducibility. 
In contrast, \emph{external library import} is significantly associated with issue reproducibility. In practice, native libraries can be imported easily following IDE recommendations, whereas importing external libraries can be more complex and time-consuming. Additionally, features like exception handling and the presence of a method have minimal influence on issue reproducibility in both SHAP and statistical analyses.

\subsection{Generalizability of our extracted features.}
\label{subsec:feature-generalizability}

We randomly select 100 questions related to C\# to check the effectiveness of our extracted features to predict issue reproducibility in similar languages like Java. We manually investigate the reproducibility of issues reported in the questions and record their status (reproducible/irreproducible). We then extract the features, train and test ML models, and evaluate their performances. These steps are discussed below.

\smallskip\noindent\textbf{Dataset Preparation.}
We followed Mondal et al.'s \citep{mondal2022reproducibility} approach in selecting and manually analyzing the dataset. We collected SO questions related to C\# posted on or before May 2023 using StackExchange API \citep{datadumpapi}. There were 74,890 questions with \emph{$<$c\#$>$} tag.
We only considered questions with code snippets to reproduce reported issues. Out of 74,890 C\# questions, 60,458 (80.7\%) included at least one line of code. We extract code snippets from questions using \texttt{<code>} tags under \texttt{<pre>}. However, a question can have multiple code snippets, which we combined (if required) or investigated separately to reproduce the issue.

Questions containing code snippets might not always discuss an issue. Programmers might seek general help or more efficient code, but we target only questions that discuss at least one programming-related issue and have at least one code snippet.
We thus look for the keywords (e.g., error, issue, exception, fix) identified by Mondal et al. \citep{mondal2022reproducibility} to find questions that discuss programming-related issues. In this way, we identified a total of 25,364 C\# questions.
However, Mondal et al. \citep{mondal2022reproducibility} found that these keywords could identify target questions with 82.0\% precision, 82.4\% recall, and 82.3\% overall accuracy. We thus randomly selected 140 questions for initial analysis. Among them, we manually analyze 100 questions that genuinely discuss programming-related issues.

\smallskip\noindent\textbf{Qualitative Analysis.}
We hired two C\# developers to analyze code snippets from 100 C\# questions. One developer who took the lead responsibility has five years of experience, and the other has ten years of C\# development experience. They follow a two-step process. First, they understand the reported issues clearly and gather supporting data from the question description. Second, they attempt to reproduce the reported issues using the code snippets and supporting data. Finally, they marked issues as reproducible or irreproducible based on their findings. The second developer investigated issues only when the first developer could not reproduce them. Issues were labeled as \emph{irreproducible} when both developers failed to reproduce them. They were able to reproduce 69 issues, while 31 could not be reproduced.

\smallskip\noindent\textbf{Environment Setup.}
We use \emph{Microsoft Visual Studio 2022 Community Edition (version 17.3.6)}\footnote{\url{https://visualstudio.microsoft.com/vs/community}} for executing code snippets and reproducing programming issues. 
For parsing C\# code snippets, we utilize \emph{Roslyn}, a tool provided by Microsoft that offers rich APIs (e.g., CSharpSyntaxTree.ParseText) for parsing the abstract syntax tree from the C\# code \citep{mondal2023subjectivity}. We use the built-in compiler \emph{Microsoft C++ Compiler (MSVC)} to compile C\# code snippets and MySQL Workbench\footnote{\url{https://www.mysql.com/products/workbench}} 8.0 as our example database for this study. We use a desktop computer with a 64-bit Windows 10 Operating System and 8GB primary memory (i.e., RAM). About 50\%-60\% of RAM was allocated (by default) for the Visual Studio IDE.

\begin{table}[htb]
\centering
\caption{Model performances of predicting reproducibility status of C\# code snippets}
\label{table:model-validation}
\resizebox{5in}{!}{%
\begin{tabular}{l|l|c|c|c|c}
\toprule
\multicolumn{1}{c|}{\textbf{Technique}} & \multicolumn{1}{c|}{\textbf{Category}} & \textbf{Precision} & \textbf{Recall} & \textbf{F1-Score} & \textbf{Accuracy} \\ \midrule

\multirow{2}{*}{\textbf{RF}} &  \textbf{Reproducible}     & 78.6\%  & 71.0\%  & 74.6\%  & \multirow{2}{*}{75.8\%}     \\ \cmidrule{2-5}
                             & \textbf{Irreproducible}    & 73.5\%  & 80.7\%  & 74.6\%  &                   \\ \midrule
                             
\multirow{2}{*}{\textbf{XGBoost}} & \textbf{Reproducible}     & 79.3\%  & 74.2\%  & 76.7\%  & \multirow{2}{*}{77.4\%}     \\ \cmidrule{2-5}
                                  & \textbf{Irreproducible}   & 75.8\%  & 80.7\%  & 78.1\%  &                         \\ \bottomrule

\end{tabular}
}
\end{table}

\smallskip\noindent\textbf{Model Construction and Performance Analysis.}
We extract the nine code-related features (Section \ref{subsec:feature-extraction}) from C\# code snippets. We select two top-performing models (i.e., RF and XGBoost) to experiment with C\# samples. Using a consistent setup (Section \ref{subsec:performance-analysis}), we apply 10-fold cross-validation and address class imbalance with SMOTE \citep{chawla2002smote}. Performance evaluation includes precision, recall, F1-score, and classification accuracy metrics.

Table \ref{table:model-validation} shows the experimental results. Both models demonstrate promising performance in predicting the reproducibility status of issues in C\# questions. Specifically, they achieve approximately 79\% precision, 71\% -- 74.2\% recall, 74.6\% -- 76.7\% F1-score, and 75.8\% -- 77.4\% overall accuracy. These results align closely with our earlier experiments using Java samples (Section \ref{subsec:model-performance}), indicating that the extracted features effectively predict reproducibility status across statically typed languages like Java and C\#.

\begin{table}[!htb]
\centering
	\caption{Editing actions to make the code snippets capable of reproducing the question issues.}
	\label{table:editing-actions}
 	\resizebox{5.7in}{!}{%
    \begin{tabular}{p{7em}|p{15em}|p{14em}|c} \toprule
    
    \textbf{Question No.}        &  \multicolumn{1}{c|}{\textbf{Reproducibility Challenges}} & \multicolumn{1}{c|}{\textbf{Action Detail}} & \textbf{No Action}  \\ \midrule
    
    \multirow{5}{*}{01 (see Fig. \ref{fig:reproducibility-challenge-1})} & (1) Class/Interface/Method not found, and (2) Important part of code missing & Addition of demo classes and methods (62\%), object creation, identifier declaration and initialization (20\%), invocation of methods (4\%) & \multirow{5}{*}{36\%} \\ \midrule
    
    \multirow{6}{*}{02 (see Fig. \ref{fig:reproducibility-challenge-2})} & (1) External library not found, (2) Identifier/Object type not found, and (3) Too short code snippet & Addition of demo classes and methods (26.5\%), the inclusion of native and external libraries (14.3\%), object creation, identifier declaration and initialization (12.2\%), other (8.2\%)   & \multirow{6}{*}{61.2\%} \\ \midrule
    
    \multirow{3}{*}{03 (see Fig. \ref{fig:reproducibility-challenge-3})} & (1) Class/Interface/Method not found, and (2) Database/File/UI dependency & Addition of demo classes and methods (28\%), inclusion of native libraries (12\%), other (4\%)  & \multirow{3}{*}{68\%}  \\ \midrule
    
    \multirow{5}{*}{04 (see Fig. \ref{fig:reproducibility-challenge-4})} & Outdated code & Addition of demo classes and methods (20.5\%), inclusion of native and external libraries (20.5\%), suggest new code (10.3\%), other (5.1\%)  & \multirow{5}{*}{53.8\%} \\ \bottomrule 
                     
    \end{tabular}
    }
\end{table}

\section{Discussion}
\label{sec:discussion}

This section discusses how SO users edit code snippets to reproduce issues aiming to provide automatic code completion insight. We also briefly discuss the potential use of LLMs for these tasks.

We analyze participants' efforts in editing code to reproduce question issues. Specifically, we examine their strategies for addressing reproducibility challenges of the code snippets (Figures \ref{fig:reproducibility-challenge-1} through \ref{fig:reproducibility-challenge-4}). Survey participants (Section \ref{subsec:survey}) were asked to submit modified code snippets, which we manually labeled the editing actions. Table \ref{table:editing-actions} summarizes the participants' code editing actions to reproduce the issues. These actions include - (a) adding demo classes and methods, (b) importing necessary native and external libraries, (c) creating objects, (d) declaring and initializing identifiers, and (e) updating outdated code.

The participants' primary objective in these editing actions is to ensure that the code snippets are compilable and executable. For instance, they observed method invocations without corresponding definitions and thus proceeded to add these definitions. The code snippets were modified based on the hints from question descriptions or guessing. However, our analysis reveals that 36\% -- 68\% of participants chose not to modify the code snippets, possibly due to perceived insufficiency or difficulty reproducing the issues accurately. For example, one participant expressed, \emph{``I did not try to reproduce it because it needs much guessing and it may differ much from the original code''}.

In recent years, natural language processing and code completion tasks have advanced significantly with the emergence of large language models (LLMs) such as GPT. LLMs can be leveraged to (a) estimate the reproducibility status and (b) identify crucial missing components in code snippets, which are vital for reproducing issues. LLMs can also be utilized to complete code snippets to make them compilable/executable for reproducing issues.

\section{Threats to Validity}
\label{threattovalidity}

Threats to internal validity relate to experimental errors and biases \citep{tian2014automated}. Our key reproducibility challenges were derived from a qualitative study by Mondal et al. \citep{Mondal-SOIssueReproducability-MSR2019}, which could be a source of subjective bias. However, the challenges were validated by 53 developers with an agreement level of about 90\% on average. Moreover, the difference between agreement and disagreement on reproducibility challenges is quite large (about 77\%) and statistically significant.

Threats to external validity relate to the generalizability of our findings \citep{tian2014automated}. Our challenges were derived by analyzing questions related to Java programming problems. Thus, the impacts and tool design requirements can apply to statically typed, compiled programming languages such as C++ and C\#. Nonetheless, replicating our study using different languages (e.g., dynamically typed languages like Python) may prove fruitful.

Our extracted features may not predict the reproducibility status of programming languages other than Java. To address this, we manually reproduced issues of 100 C\# questions, extracted features, and built ML models. The results were promising, confirming our features' generalizability. However, due to their flexibility, these features may not work well for dynamically typed languages like Python.

We only used data from SO that might cause generalizability threats. We could not validate our findings with other technical Q\&A forums because their data is not readily available, and their question and code structures are not well-defined for automatic data collection. However, manually collecting a large number of data samples is impractical. Future research using data from other forums could confirm our results.

Our online survey may have a selection bias. We used both a snowball approach and an open circular for participant selection. Responses were collected anonymously to encourage honesty. Participants ranged from novice to experienced (see Fig. \ref{fig:java-experience}) and included mainly software developers, along with other related professions (see Fig. \ref{fig:profession}). This diversity adds validity to our findings, and any individual biases are mitigated by the large sample size of 53 users.

\section{Conclusions}
\label{conclusion}

Developers post thousands of questions on SO to resolve their code-level problems and include example code snippets to support the problem descriptions. However, these code snippets cannot always reproduce the question issues due to several unmet challenges, preventing prompt and appropriate solutions. A previous study produced a catalog of reproducibility challenges, but practitioners did not validate them. In this study, we attempt to understand practitioners' perspectives on these challenges and (2) develop five ML models to predict the reproducibility status of issues.
Our study findings are five-fold. 
\emph{First,} about 90\% of developers agree with challenge catalog.
\emph{Second,} ``missing an important part of code'' most prevents questions from receiving answers. 
\emph{Third,} a tool to analyze code snippets and identify reproducibility issues could help users improve their questions.
\emph{Fourth,} our ML model can predict issue reproducibility with 84.5\% precision, 83.0\% recall, 82.8\% F1-score \& 82.8\% overall accuracy.
\emph{Fifth,} we systematically interpret the results of the ML model.

In the future, we plan to integrate tool support (e.g., browser plugin) with the SO's question submission system to assist question submitters in improving their code snippets to promote reproducibility. LLMs can be leveraged to analyze code snippets and offer immediate feedback and suggestions for improving them and their reproducibility.

\vspace{1mm}
\textbf{Acknowledgment:} This research is supported in part by the Natural Sciences and Engineering Research Council of Canada (NSERC) Discovery Grants program and by the industry-stream NSERC CREATE in Software Analytics Research (SOAR). This research is also supported in-part by two Canada First Research Excellence Funds (CFREFs) grants coordinated by the Global Institute for Food Security (GIFS) and the Global Institute for Water Security (GIWS).

\bibliographystyle{elsarticle-num-names} 
\bibliography{bibtex.bib}

\end{document}